\documentclass[twocolumn]{aastex631}
\usepackage{lmodern}
\usepackage{latexsym}
\usepackage{amsmath}
\usepackage{amssymb}
\usepackage{amsbsy}
\usepackage{amsthm}
\usepackage{amsfonts}
\usepackage{mathrsfs}
\usepackage{bm}
\usepackage{sansmath}
\usepackage{relsize}
\usepackage{caption2}
\usepackage{graphicx}
\usepackage[utf8]{inputenc} % usually not needed (loaded by default)
\usepackage[T1]{fontenc}
\usepackage{epstopdf}
\usepackage{esdiff}
\usepackage[T1]{fontenc}
\usepackage{ae,aecompl}
\shorttitle{Outflow in SRAF}
\shortauthors{Ranjbar \& Abbassi}
\graphicspath{{./}{figures/}}

\begin{document}

\title{Slowly Rotating Accretion Flow Around SMBH in Elliptical Galaxy: Case With Outflow}

\email{abbassi@um.ac.ir, razieranjbar69@gmail.com}

\author[0000-0003-1488-4890]{Razieh Ranjbar}
\affiliation{Department of Physics, Faculty of Science, Ferdowsi University of Mashhad, Mashhad, Iran}

\author[0000-0003-0428-2140]{Shahram Abbassi}
\affiliation{Department of Physics, Faculty of Science, Ferdowsi University of Mashhad, Mashhad, Iran}




\begin{abstract}

Observational evidence and many numerical simulations show the existence of wind (i.e., uncollimated outflow) in accretion
systems of the elliptical galaxy center. One of the primary aims of this study is to investigate the
solutions of slowly rotating accretion flows around the supermassive black hole with outflow. This paper presents
two distinct physical regions: supersonic and subsonic, that extend from the outer boundary
to the black hole. In our numerical solution, the outer boundary is chosen beyond the Bondi radius. Due to strong gravity, we ignore outflow (i.e., $ s = 0 $) in the inner region (within $ \sim 10 r_s $). The radial velocity of the flow at the outer region is significantly increased due to the presence of the outflow. Compared to previous works, the one accretion mode— namely, slowly rotating case, which corresponds to low accretion rates that have general wind output is carefully described, and the effect of galaxy potential and the feedback effects by the wind in this mode are taken into account. Since the power-law form of the mass accretion rate is mathematically compatible with our equations, we consider a radius-dependent mass accretion rate ($ \dot{M}_{in} \propto r^s $), where $s$ is a free parameter and shows the intensity of outflow. There is an unknown mechanism for removing the mass, angular momentum, and energy by outflows in this study. The effects of the outflow appear well on the outer edge of the flow. 

\end{abstract}


\section{Introduction} \label{sec:intro}

A majority of cool-core clusters have massive black
holes in their central regions, which produce persistent
relativistic jets. It is important to point out that these
jets are highly interactive with the ambient medium,
and they inject a large amount of energy into the surrounding gas, which is key to AGN feedback. M87 is a supergiant elliptical galaxy near the center of the Virgo cluster that serves as a unique laboratory for studying AGN jets and their formation and acceleration thanks to its proximity with a distance of $16.7$ $\rm M \rm pc$ and its extremely massive black hole with an average value of the mass of $M_{\rm BH}= 6.5 \times 10^9 M_{\odot}$ (The mass measurements that have been carried out with Event Horizon Telescope data) (\cite{2019ApJ...875L...6E}). Active galactic nuclei (AGNs’) feedback plays a crucial role in the evolution of its host galaxy (\cite{2012ARA&A..50..455F}; \cite{2015ARA&A..53..115K}). Over the past 20 years, there has been tremendous growth in the literature relating to AGN
feedback (\cite{2018NatAs...2..198H}). The outputs of the AGN, such as
radiation, wind, and jet, interact with the interstellar
medium (ISM) in the host galaxy and change the density and temperature of the ISM. It follows that star formation activity will be affected, and accordingly, galaxy
evolution will be affected as well. For evaluating the effects of AGN feedback, the value of accretion rate and
model of accretion physics plays a critical role since they
determine the magnitude of AGN outputs.

In order to study AGN feedback, one of the most important quantities is the mass accretion rate of the BH, which has a great deal to do with AGN power. There is a widely held belief that much of the energy behind astronomical objects is due to mass accretion onto black holes, as seen in active galactic nuclei (AGNs), X-ray binaries, and even gamma-ray bursts (\cite{2002apa..book.....F}). It may be appropriate to use the spherical Bondi model to describe galaxies' evolution since the central AGN is often found in the region with the densest atmosphere within the cluster. (\cite{1952MNRAS.112..195B}; \cite{2003ApJ...591..891B}; \cite{2003ApJ...582..133D}). AGNs can receive sufficient fuel from the Bondi accretion rate based on Chandra's observations of M87 \cite{2003ApJ...582..133D}. In addition, the Bondi model is an affordable option for modeling hot accretion flows. The Bondi model is undoubtedly one of the best transonic models available today. But a spherically symmetric Bondi accretion cannot occur in nature and black hole accretion processes are almost always associated with rotating gas flows, so a modified Bondi model is needed. \cite{2011MNRAS.415.3721N}, hereafter \url{NF11}, discussed low-angular momentum accretion and added a slow rotation to Bondi accretion flows. As an intermediate case between Bondi and advection-dominated accretion flows (ADAF), this model was applied to Sgr A* and M87. In recent years, Bondi solutions and slowly rotating accretion flow have been developed by taking into account galaxy gravitational potential (\cite{2022MNRAS.516.3984R} hereafter \url{R22}; \cite{2016MNRAS.460.1188K}; \cite{2017ApJ...848...29C}; \cite{2018ApJ...868...91C}; \cite{2019MNRAS.489.3870S}). Early analytical work assumed that mass accretion rates were independent of radius, $ \dot{M}(r) = constant $, so in this case, radial density profiles satisfy $ \rho \propto r^{-3/2} $ (e.g., \cite{1994ApJ...428L..13N}).
The dynamics of the accretion flow have been studied extensively in hydrodynamic (HD) and magnetohydrodynamic (MHD) numerical simulations (e.g., \cite{2000ApJ...537L..27I};
\cite{1999MNRAS.310.1002S};
\cite{2003ApJ...582...69P};
\cite{2012MNRAS.423.3083M};
Sgr A* and NGC 3115 are two of the best targets of Chandra's observations that show flat density profiles with $\rho \propto r^{-1}$ inside the Bondi radius (\cite{2013Sci...341..981W}). \cite{2015MNRAS.451..588R} also reported a similar outcome, with a density profile of $\rho \propto r^{-1}$ observed for M87. Regarding recent numerical simulations, only a fraction of the Bondi rate is able to reach the SMBH due to outflows originating from the accretion flow. (e.g. \cite{2012ApJ...761..129Y}; \cite{2013ApJ...767..105L}). It seems that objects created by accretion, such as galaxies and stars, have generated outflows that have significantly impacted the surrounding environment. There is also a belief that the outflow plays a more significant role in suppressing star formation and the BH accretion rate (\cite{2018ApJ...857..121Y}). It is therefore worthwhile to consider outflow in Bondi-like accretion flow physics.

Recent observations and simulations suggest that outflows/winds play a significant role in accretion systems. One of the most important findings in recent years in hot accretion flows has been the discovery of the strong outflow/wind that originates from the accretion flow (\cite{2012MNRAS.426.3241N}; \cite{2012ApJ...761..130Y}; \cite{2013ApJ...767..105L}). 
Due to the highly ionized nature and absence of an absorption line of a hot accretion flow, it is difficult for us to observe wind directly from it \cite{2019ApJ...882...55B}.
The presence of wind in hot accretion flows is evident from indirect observations. Observation of the 3 million seconds of Chandra from the supermassive black hole in the center of our galaxy, Sgr A*, confirmed this theoretical prediction recently (\cite{2013Sci...341..981W}).
It is interesting to find that wind is not only a key factor in accretion physics but also plays a significant role in the feedback of AGNs as well (e.g., \cite{2010ApJ...722..642O}; \cite{2014ApJ...789..150G}). Observations of LLAGNs  (e.g., \cite{2012ApJ...753...75C}; \cite{2013Sci...341..981W}; \cite{2016Natur.533..504C}), as well as black hole X-ray binaries (\cite{2016ApJ...830L...5H}), have shown that hot accretion flows are influenced by winds. Also, the existence of outflows in the radiatively inefficient accretion flows is confirmed by many observations (e.g., \cite{2013Sci...341..981W}, \cite{2019MNRAS.483.5614M}, \cite{2019ApJ...879L...4M}). Therefore, the numerical simulation result is confirmed by observations. In accordance with numerical simulations and analytical models of the hot accretion flow, the accretion rate decreases with decreasing radius, so most of the gas is lost between the Bondi radius and the event horizon (e.g., \cite{1999MNRAS.303..309I};
\cite{1999MNRAS.310.1002S}; \cite{2010MNRAS.408.1051Y}; \cite{2012MNRAS.420.2912B}). Many analytical models have been developed in recent years to investigate how hydrodynamical wind affects the structure of ADAFs (\cite{2010MNRAS.409.1113A}; \cite{2014MNRAS.437.3112M}; \cite{2016MNRAS.456...71G}). The outflow is particularly crucial since it carries mass, momentum, and energy to the surrounding environment, exerting significant influence there (\cite{2022MNRAS.513.2100H}, \cite{2022PASJ...74..384B}).
Winds are believed to play a crucial role in the interaction between AGNs and their host galaxies (e.g., \cite{2010ApJ...717..708C};\cite{2010ApJ...722..642O}; \cite{2017ApJ...835...15C}; \cite{2017MNRAS.465.3291W}; \cite{2022MNRAS.513.2100H}).

\url{NF11} studied the slowly rotating
accretion flow in the region very close to the black hole
$r_{s}$ ($r_{s}$ is Schwarzschild radius) to the region beyond the
Bondi radius. We recalculate \url{NF11} solutions for accretion flow with small angular momentum by including outflow effects. We find that the real accretion rate
onto the black hole deviates from the value calculated
by \url{NF11} significantly. In this paper, we
study the slowly rotating accretion flow in the 
region from $0.01$ of $r_{\rm s}$ to the outer boundary. The outer
boundary is quite large, almost three times larger than the
Bondi radius. This paper differs from that of \url{NF11} in
two aspects. First, in \url{NF11}, only the black hole gravity
is taken into account. However, in the region around $10$
parsec, the gravitational force of a nuclear star is
comparable to that of the central black hole. The star's
gravity may be important and needs to be considered. In the present paper, we will take into account the gravitational potential of the galaxy's stars. Second, in \url{NF11}, the effect of outflows and significant radial exchanges of energy are suppressed. The mass conservation equation (1) of \url{NF11} explicitly ignores such outflows. In
the present work, we solve differential equations in the presence
of outflow. A general mechanism for producing outflows
is considered instead of a detailed description of power
outflow mechanisms. The presence of outflow is beneficial to study the effect of AGN feedback on galaxy formation and evolution. 
It is complex to consider the interactions between the outflowing energy, mass, and momentum with the ambient fluids, and we postpone this discussion to the next article in the series.
The rest of the paper is planned
accordingly: In section \ref{thickness}, we investigate two cases. In the first case, we have compared our results with those of \url{BY19}, and in the second case, we have investigated how outflow affects the thickness of flow by considering a more realistic state, where $ H \neq r $. Finally, in section \ref{sec.AGN}, we investigate the AGN feedback by calculating the energy and momentum fluxes of outflow and compare it to the previous work of \cite{2018ApJ...857..121Y}. In this paper, we focus on the case without the magnetic field and convection. The paper is organized as follows.

In \S  \ref{Model}, we analyze the basic equations, and various boundary conditions and explain our numerical method. The results of the numerical solution are given in \S  \ref{3}, and finally, in  \S  \ref{sec:5}, we summarize our results.

  \section{Model} \label{Model}
\subsection{The Structure of the accretion flow in the presence of outflow}\label{equation}

In this study, we consider slowly rotating, steady, viscous accretion flows onto a non-rotating SMBH by including the role of outflow, which is commonly referred to as hot accretion flows. Such flows are nearly spherical, so the height-integrated differential
equations can describe them well. In order to simplify our model, we assume all the variables are functions only
of radius and independent of time. Our study employs a simplified approximation, $H = r$, following the approach of Narayan and Fabian (2011). This approximation corresponds to a disk that is geometrically very thick.
Under these assumptions, we can get the basic conservation equations. In the presence of outflow, the continuity equation of accretion flow is 

\begin{equation}\label{cont}
\diff{}{r}(4 \pi r^2 \rho v) + \diff{\dot{M}_{\rm w}}{r} = 0
\end{equation}

where $ \rho $ is the density of the gas, $ v $ is the radial inflow velocity of gas ($ v < 0 $), $ \dot{M}_{\rm w} $ is the mass loss rate by outflow. It can be concluded that the mass accretion rate varies with radius according to this equation.

The mass-loss rate $ \dot{M}_{\rm w} $ is adopted from \cite{1999MNRAS.309..409K} hereafter \url{K99}: 

\begin{equation}
\dot{M}_{\rm w} (R) = \int_{r_{\rm c}}^r \dot{m}_{\rm w}(r) dr,
\end{equation}

where $ r_c $ denotes the sonic point radius and $ \dot{m}_w $ is the mass loss rate per unit area from each flow face.
Whenever mass is carried away by outflow, the mass accretion rate decreases inward. The mass accretion rate is expressed using the power-law form for several reasons below. The first benefit is that makes solving differential equations simple. On the other hand, all lunching wind mechanisms can be described with the \url{K99} wind model, which is a general and parametric model. Lastly, the power-law form is mathematically compatible with our governing differential equations. We adopt the power-low form for $ \dot{M}_{\rm in} $ as follows

\begin{equation}\label{cont2}
\dot{M}_{\rm in} =  - 4 \pi r^2 \rho v = \dot{M}_{\rm out} (\frac{r}{r_{\rm out}})^{s}, \quad  for  \quad  r_{\rm c} < r < r_{\rm out}
\end{equation}

where $s$ is the mass loss power-law index and represents the strength of the outflow. Also, $r_{\rm out}$ and $ \dot{M}_{\rm out} $ are the radius and the mass accretion rate in
the outer boundary, respectively. A simulation paper
by \cite{2005ApJ...628..368O} demonstrates that s is not a
constant in the flow, but its average value can be approximated around 1 (\cite{2018ApJ...860..114K}). The analytical approach has limitations, so we considered ”$ s 
$”
as a constant and a parameter in a particular solution
($ 0 < s < 1 $). The mass accretion rate at the black hole
horizon is 

\begin{equation}
\dot{M}_{\rm BH} = \dot{M}_{\rm out} (\frac{r_{\rm c}}{r_{\rm out}})^s
\end{equation}

In this work, we assume $ \dot{M}_{\rm BH} $ equal to $ \dot{M}_{\rm c} $, the mass
accretion rate in the sonic point. Actually, the power-law variation of $ \dot{M}_{\rm in} $ with $ r $ is not likely to continue all the way down to $ r_s $, and it will probably cease somewhere near $r_{\rm in}$ of order ten (or even tens of) $ r_s $ (\cite{2014ARA&A..52..529Y}). Since in the present study, the sonic
radius is constrained to approximately $10 r_s$, we assume
that the power-law variation of $ \dot{M}_{\rm in} $ will continue to $ r_{\rm c} $.
With the integration of equation \ref{cont}, the mass flux of outflow gives 

\begin{equation}\label{wind}
\dot{M}_{\rm w} (r) = \dot{M}_{\rm in}(r) -  \dot{M}_{\rm BH}
\end{equation}

The second term of the right side is the net mass accretion rate that is accreted into SMBH. It is clear that both $\dot{M}_{in}$ and $\dot{M}_{out}$ are a power-law function of $r$. However, the difference between the two ($\dot{M}_{BH}$) remains constant across the radius similar to those simulations (\cite{1999MNRAS.310.1002S}). This difference is approximately equal to the local mass accretion rate expected for a viscous flow with low angular momentum. This net mass accretion rate is swallowed by the black hole. According to Equation \ref{wind}, we have:

\begin{equation}
d \dot{M}_{\rm in} / dr = d \dot{M}_{\rm w} / dr 
\end{equation}

On the other hand, it has been found that the gravitational influence of the galaxy's stars should not be ignored at the parsec scale
(\cite{2016ApJ...818...83B}; \cite{2018MNRAS.478.2887Y}). In calculating the total gravity, we consider the gravitational force of
the stars in the galaxy since our computational scale covers sub-parsec and parsec scales. 
The total gravitational potential is as follows:

\begin{equation}
\Phi = \phi_{\rm BH} + \phi_{\rm galaxy}
\end{equation}

where $\phi_{\rm BH}$ is the potential of a black hole to emulate the general relativistic effects of a Schwarzschild black hole, we use the gravitational potential \cite{1980A&A....88...23P}. 
\begin{equation}
\phi_{\rm BH} = -\frac{GM}{r-r_{\rm s}}
\end{equation}
where $ M $ is the mass and $ r_{\rm s} \equiv 2GM / c^2 = 2r_{\rm g} $ denotes the Schwarzschild radius of the non-rotating black hole. The self-gravity of the accretion flow is neglected. And $\phi_{\rm galaxy}$ is the galaxy potential.
\begin{equation}
\phi_{\rm galaxy} = \sigma^2 ln r + C
\end{equation}
where $C$ is a constant and $\sigma$ is the dispersion velocity of stars (\cite{2016ApJ...818...83B}). It is common for elliptical galaxies with a central black hole mass $ M_{\rm BH} = 10^8 M_{\odot}$ to have a stellar dispersion velocity of $150-250$ $km s^{-1}$.
Similarly to \url{R22}, we consider the velocity dispersion of stars to be a constant of radius, which is $\sigma = 200 km/s $ (\cite{2013ARA&A..51..511K}).

The radial momentum equation is

\begin{equation}\label{radialmom}
v \diff{v}{r} = - (\Omega_{\rm K}^2 - \Omega^2) r - \frac{1}{\rho} \diff{}{r} (\rho c_{\rm s}^2) -\frac{\sigma^2}{r}, 
\end{equation}

where $ \Omega_{\rm k} $ is Keplerian angular velocity. We denote the isothermal sound speed  by $ c_{\rm s} = (P/\rho)^{1/2}  $, and $ P $ is the total pressure of the gas.

In our model, it is assumed that matter ejected from flow at radius $ r $ carries away specific angular momentum $  (l r)^2 \Omega $. Where $ l $ is a dimensionless parameter representing the amount of angular momentum carried away by outflow, and $ \Omega $ is the angular velocity in the flow at radius $ r $. Owing to the presence of the outflow, the angular momentum equation takes the form

\begin{equation}\label{mom}
\diff{}{r}(4\pi r^4 \rho v \Omega) = \diff{}{r}(4 \pi \nu \rho r^4 \diff{\Omega}{r}) - (l r)^2 \Omega \diff{\dot{M}_{\rm w}}{r}
\end{equation}

Outflows with $l = 0$ correspond to non-rotating outflow, so they cannot extract angular momentum, and the flow only loses mass. While  $l > 0$ indicates outflowing materials that carry away the angular momentum (\cite{1999MNRAS.309..409K}; \cite{2013ApJ...765...96A}; \cite{2016MNRAS.456...71G}; \cite{2019ApJ...887..256H}). Furthermore, $ 0 < l < 1 $ can be classified as a family of outflows that carries away less angular momentum than the outflow material had before leaving the accretion flow. Models  $ l > 1 $ also effectively describe centrifugally driven magnetic outflows, which can remove a large amount of angular momentum from flow \cite{1982MNRAS.199..883B}.
We note that the hydrodynamical outflows are in stark contrast with the magnetically driven outflows, in which the gas may move along the field lines corotating with the flow to a distance far from the flow surface, and therefore a substantial fraction of the flow angular momentum may be removed through the magnetically driven outflows (e.g., \cite{1994MNRAS.268.1010L}; \cite{2002A&A...385..289C}, \url{2013}). In the second term on the right-hand side of equation \ref{mom}, the outflow provides an angular momentum sink. In this model, a simple approach can be
applied to a wide range of outflow models; for more information, see \url{K99}, where $ \nu $ is the kinematic coefficient of viscosity that is defined by

\begin{equation}\label{nu}
\nu = \alpha c_{\rm s} r
\end{equation}

As mentioned in \cite{1973A&A....24..337S}, $ \alpha $ is the viscosity parameter and is usually assumed to be a constant. Narayan and Fabian (2011) integrated simply the angular momentum equation (equation (7) of \url{NF11}). But you can see in equation \ref{mom} above the additional term due to outflow makes it much more difficult to integrate self-consistently. We will drive equation \ref{intmom} from equation \ref{mom}, by integration and using an approximate of $ \Omega \propto r^{-3/2} $ for simplicity. 
This approximation in this study is not far from reality, since angular velocity is a coefficient of Keplerian angular velocity or actually sub-Keplerian. Integrating Equation \ref{mom} yields 

\begin{equation}\label{intmom}
 \diff{\Omega}{r} = \frac{v}{\alpha c_{\rm s} r^3} (\Omega r^2 (1- \frac{s  l^2}{s+1/2}) -j)
\end{equation}

Where the integration constant $ j $ represents the specific angular momentum per unit mass, which is swallowed by the black hole.
 It is assumed that matter ejected from the accretion flow carries away energy in addition to mass and angular momentum. To overcome the energy of the accretion flow, the outflow material must be supplied with energy at a rate of $\frac{1}{2} \eta \dot{m}_w v_k^2$ per unit area. The fraction $\eta$ of the outflow's energy at infinity is derived from the accretion energy dissipated, while $\dot{m}_w$ represents the mass loss rate per unit area from each flow face. This equation is used to model the energy output of the accretion flow and the resulting outflow in our analysis. Eventually, the energy equation reads 

\begin{equation} \label{energy}
\frac{2 r \rho v}{(\gamma - 1)} \frac{d c_{\rm s}^2}{dr} - 2 r c_{\rm s}^2 v \frac{d \rho}{dr} = 2 r \rho f \nu r^{2} \left( \frac{d \Omega}{dr} \right)^2 - \frac{1}{2} \eta \dot{m}_{\rm w} v_{\rm k}^2
\end{equation}

The adiabatic index of the gas is $ \gamma $ which, in our work, equals $ 5/3 $. $\eta$ is a free and dimensionless parameter in our model. In accordance with the energy loss by outflow, the parameter $\eta$ may change (\url{K99}). In most cases, we are interested in a radiatively inefficient flow in which radiative cooling is neglectable. Hence, the first term on the left-hand side of equation \ref{energy} represents the viscous heating rate per unit volume, and the second term on the left-hand side of it is losing energy by outflow (\url{K99}). In order to measure the degree to which the flow is advective, we use the parameter $ f $ (\cite{1994ApJ...428L..13N}). Generally, for radiative-cooling-dominated flows $ f \sim 0$, such as SSDs or SLE disks, while for advection-dominated flows $ f \sim 1$. Hence, in this paper, we assume that $ f = 1 $ (full advection case). The approximation works well whenever radiative cooling is less than a few percent of the heating rate.

\subsection{The Inner Region }
A viscous, slowly rotating accretion flow with outflow is described by the equations in Subsection \ref{equation}. In the region near SMBH, accreting gas passes inside the sonic radius $r_{\rm c}$ and becomes supersonic. In some general relativistic simulations of hot accretion flows, \cite{2012ApJ...761..129Y}, \cite{2013MNRAS.434.1692B}, \cite{2013MNRAS.436.3856S}, and \cite{2012ApJ...761..130Y} have shown that the accretion rate is constant within $\sim 10 r_s$. This may be caused by the strong gravity around the black hole so $ s = 0 $, meaning no mass loss from the flow. Also, in the inner regions, gravity is greater than pressure force, and the radial velocity is considerably large (so the viscous timescale is short), which means that only a weak outflow can form \cite{2022ApJ...930..108W}; \cite{1999MNRAS.310.1002S}. Furthermore, in regions very close to the SMBH, the time-scale forming outflow is more than the accretion time scale, so we may ignore the outflow effect in this region. So we assume outflows from inside $ 10 r_{\rm s} $ are quite weak. Certainly, the simplest description that one could propose would be to imagine that no energy, mass, or angular momentum transfer from the supersonic flow to the surrounding environment. Nonetheless, in the present work, we neglect viscosity $ \alpha $ in the inner region following the similar argument given in \url{NF11}. Furthermore, it is important to note that the velocity dispersion of stars contributes significantly to the distribution of gravitational forces within the galaxy, especially when the radius exceeds $1 \rm pc$ (\cite{2018MNRAS.478.2887Y}; \url{R22}). This causes the gravitational effect of stars close to the black hole to greatly reduce or disappear completely. Thence, all of the differential equations become algebraic equations (See section 2.2 of \url{NF11} for more details).

\subsection{BOUNDARY CONDITIONS} \label{BC}

Accretion flow around the SMBH ultimately falls in the center; away from SMBH, flow is subsonic. These transonic flows are supersonic after passing the sonic radius and fall towards BH with speeds near light. Because of this, Our computational domain consists of two parts. The inner region is supersonic. The viscosity of flow in this region is low, so we consider inviscid flow and assume the inner region has inflow only and ignore outflow here. The other region is from the sonic radius to the outer boundary; in this region, we have viscous inflow/outflow fluid. The main differential equations consist of four variables. We need four boundary conditions to solve them properly. In addition, $ j $ is an eigenvalue, so we induced another constraint for adding one more boundary condition. On the other hand, the location of the sonic radius is unknown; thus, we require another boundary condition. This problem has six boundary conditions, similar to those in \url{NF11}.

Following the procedure adopted in previous works for studying transonic accretion flow around BH/compact objects (e.g., \cite{1996ApJ...464..664C}; \cite{1997ApJ...476...49N}; \cite{2003MNRAS.342..274M}), we combine equation (\ref{cont2}) with differential equations (\ref{radialmom}), (\ref{intmom}), and (\ref{energy}), so the following relation appears

\begin{equation}
\diff{ln v}{r} = \frac{\mathcal{N}}{\mathcal{D}} 
\end{equation}

\begin{multline}
\mathcal{N} = \frac{(\Omega_k^2 - \Omega^2)r}{c_s^2} +\frac{ \gamma ( s - 2 )}{r} + \frac{(\gamma - 1) \alpha r^3}{ v c_s }(\diff{\Omega}{r})^2 \\
 + (\gamma - 1) \frac{\eta}{4} \frac{s}{r c_{\rm s}^2}( r^2 \Omega_k^2) + \frac{\sigma^2}{r c_{\rm s}^2}
\end{multline}

\begin{equation}
\mathcal{D} = \gamma - \frac{v^2}{c_s^2}
\end{equation}

Since the flow in the sonic radius must be smooth, this equation gives us two boundary conditions in $ r = r_c $,

\begin{equation} 
\gamma c_{\rm s}^{2} - v^2 = 0, 
\end{equation}

\begin{multline}
( \Omega_{\rm k}^2 - \Omega^2 ) r_{\rm c}  +  \gamma (\frac{ (s - 2) c_s^2}{r_{\rm c}} + (\gamma - 1) \frac{\eta}{4} \frac{s}{r_{\rm c}} (r_{\rm c}^2 \Omega_k^2) + \\
\frac{(\gamma - 1) \alpha r_c^3 c_{\rm s}}{v}(\diff{\Omega}{r})^2 + \frac{\sigma^2}{r_c} = 0, 
\end{multline} 
 
The third boundary condition in sonic radius acquires from the angular momentum equation. Since viscosity and outflow have been ignored in the supersonic region, one may consider the specific angular momentum of the gas remains constant. Thus we have

\begin{equation}
\Omega r_{c}^{2} (1- \frac{ l^2 s}{s+1/2}) - j + \frac{2 \alpha c_{\rm s} \Omega r_c^2}{v }  = 0,
\end{equation}

The rest of the boundary conditions are in the outer boundary. In most astrophysical systems, the accretion geometry consists of two zones, the outer cold zone, as described by the thin disk model, and the inner hot zone, represented by the hot accretion mode. In previous studies, have been assumed that the flow begins in a state of a geometrically thin accretion disk \url{NKH97}. But in other cases like Sgr A* and M87, the gas starts on hot at the Bondi radius and stays hot up to the inner edge. The Bondi radius is generally considered the outer boundary of the accretion flow surrounding a supermassive black hole (M87, Sgr A*), where the thermal energy of the external surroundings is equal to its potential energy due to the black hole’s gravitational field (\cite{2014ARA&A..52..529Y}).

In the Bondi or even slowly rotating solutions, the temperature and density at the outer boundary play a crucial role and may be used as outer boundary conditions. The temperature value at the outer boundary is adopted as $ 6.5 \times 10^6 K $ which corresponds to the observed values for the interstellar medium at the nucleus of an elliptical galaxy at the center of a cool-core group or cluster of galaxies (\url{NF11}). Therefore, we have two conditions at the outer boundary

\begin{eqnarray}
c_{\rm s} = c_\mathrm{out}, \quad r = r_\mathrm{out} \nonumber \\
\rho = \rho_\mathrm{out}, \quad r = r_\mathrm{out}
\end{eqnarray}

 \begin{figure*}
\centering
\makebox[\linewidth]{%
\begin{tabular}{cc}
    \includegraphics[width=0.45\linewidth]{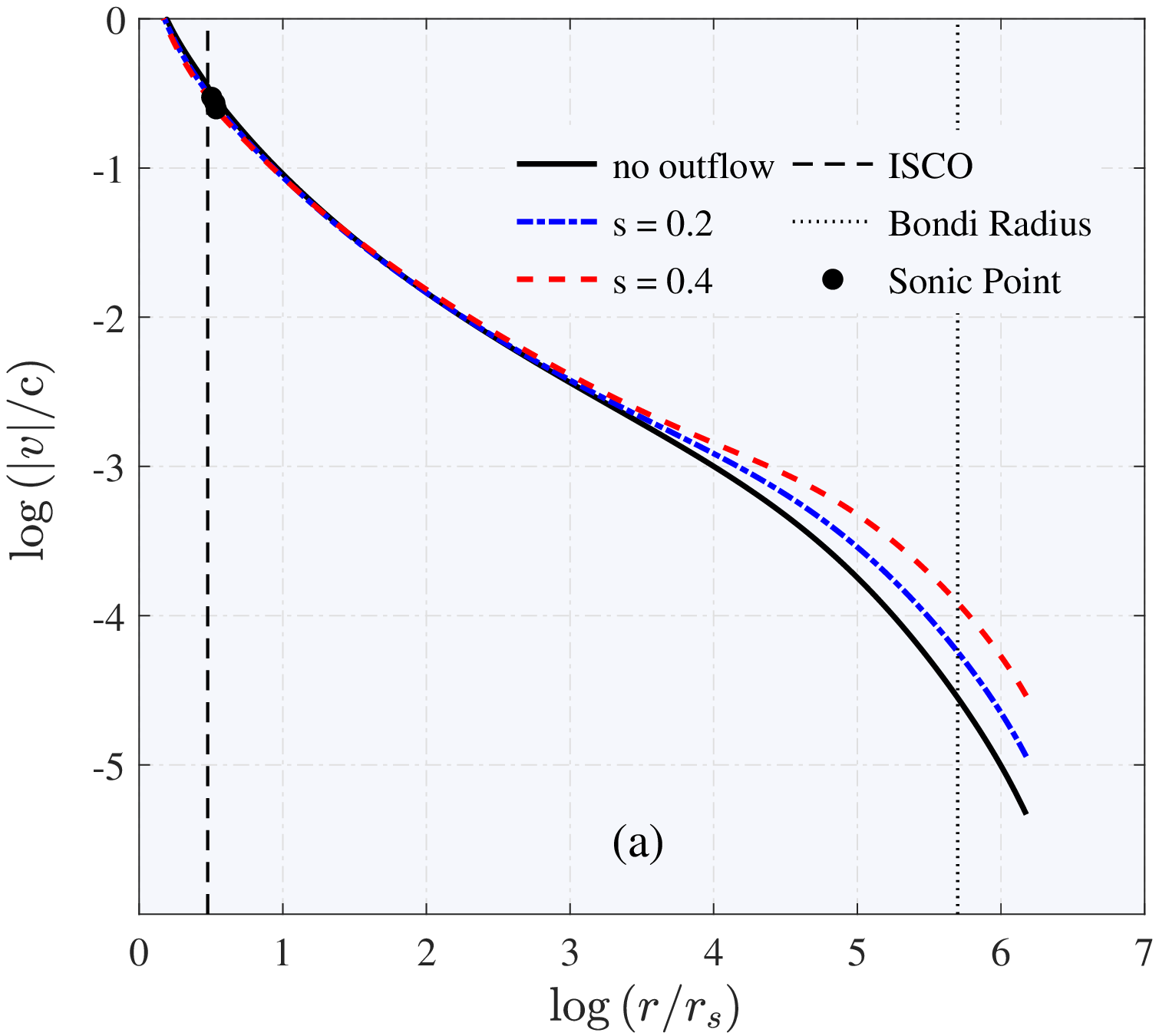} &
    \includegraphics[width=0.45\linewidth]{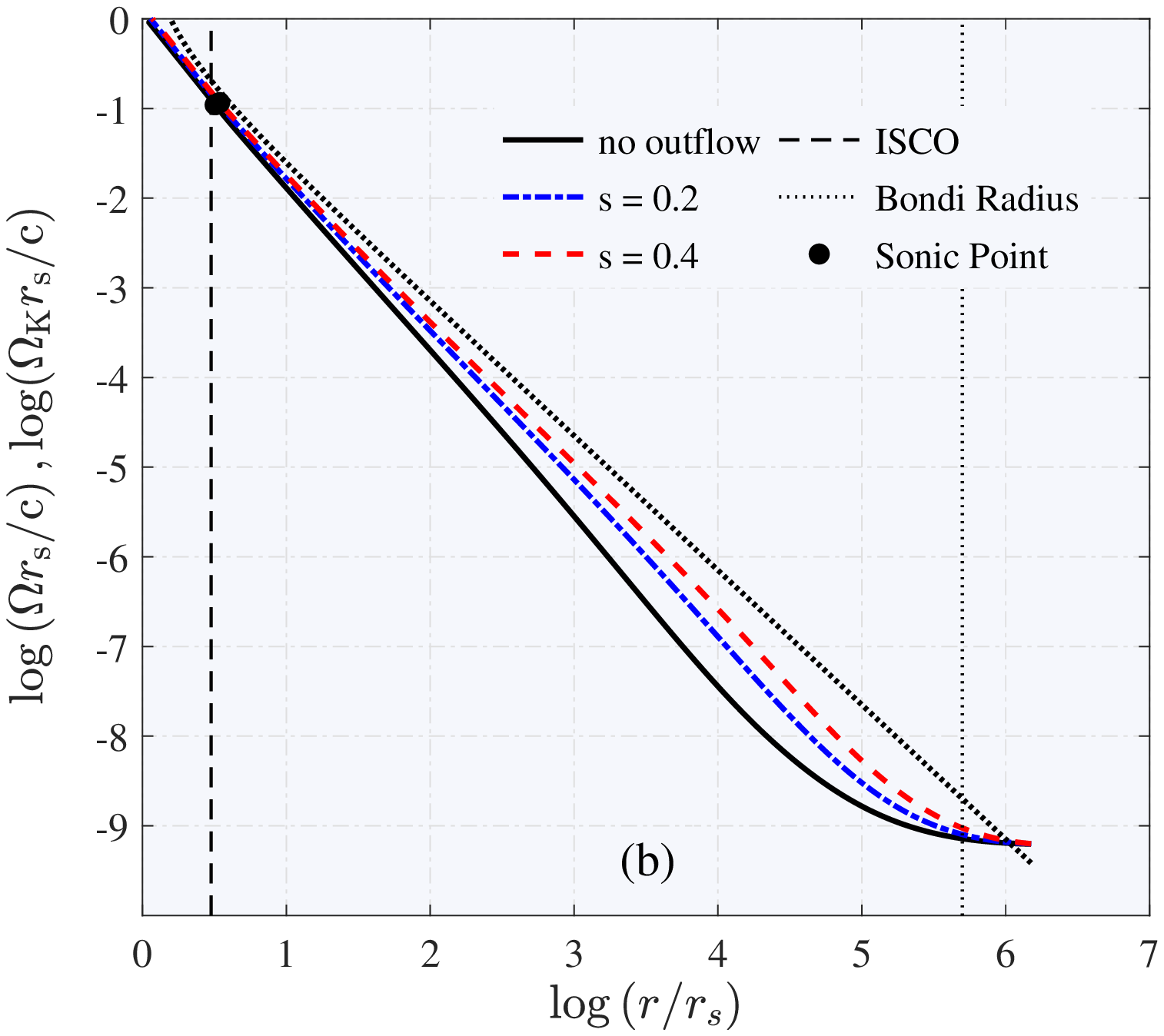} \\
    \includegraphics[width=0.45\linewidth]{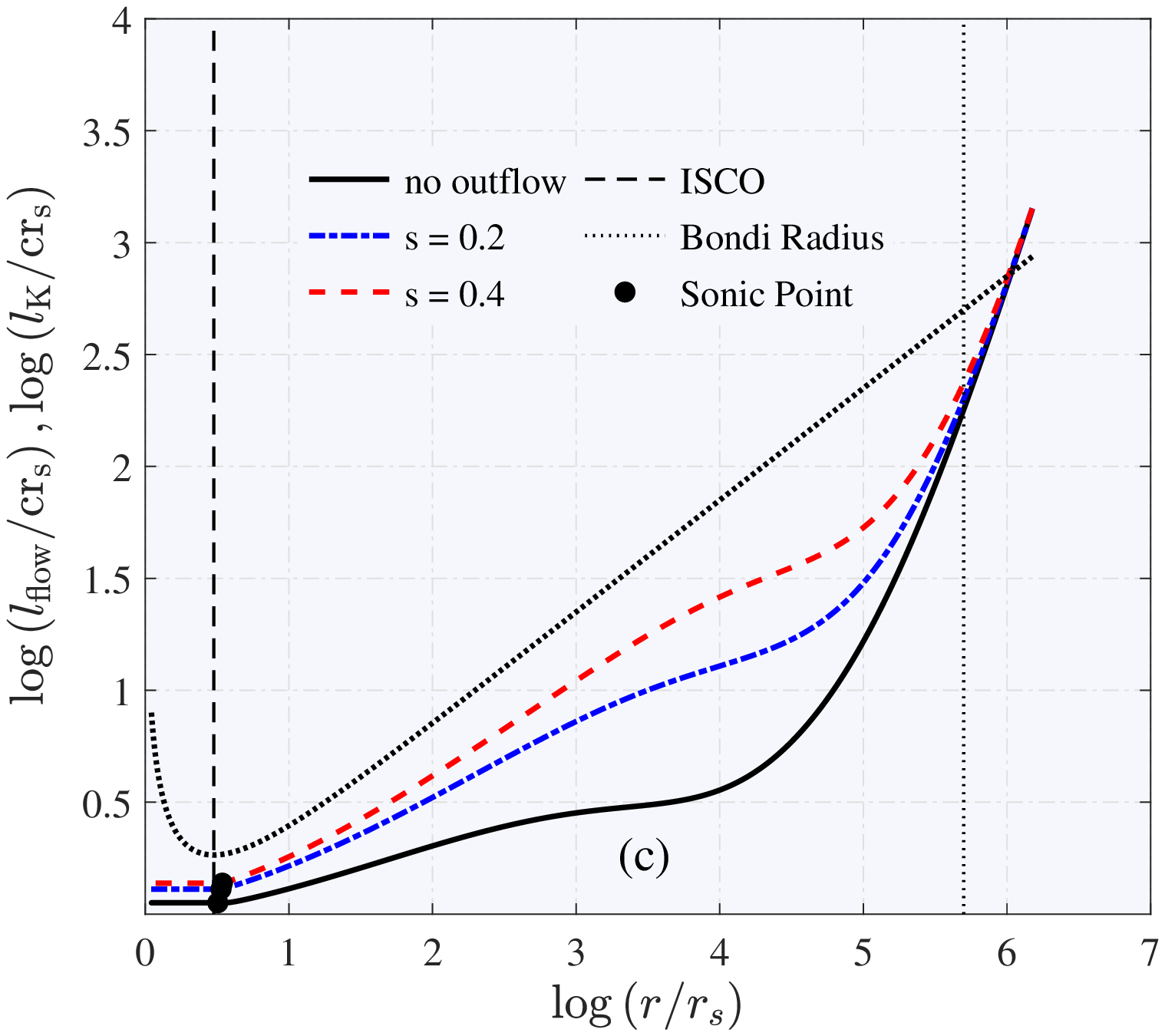} &
    \includegraphics[width=0.45\linewidth]{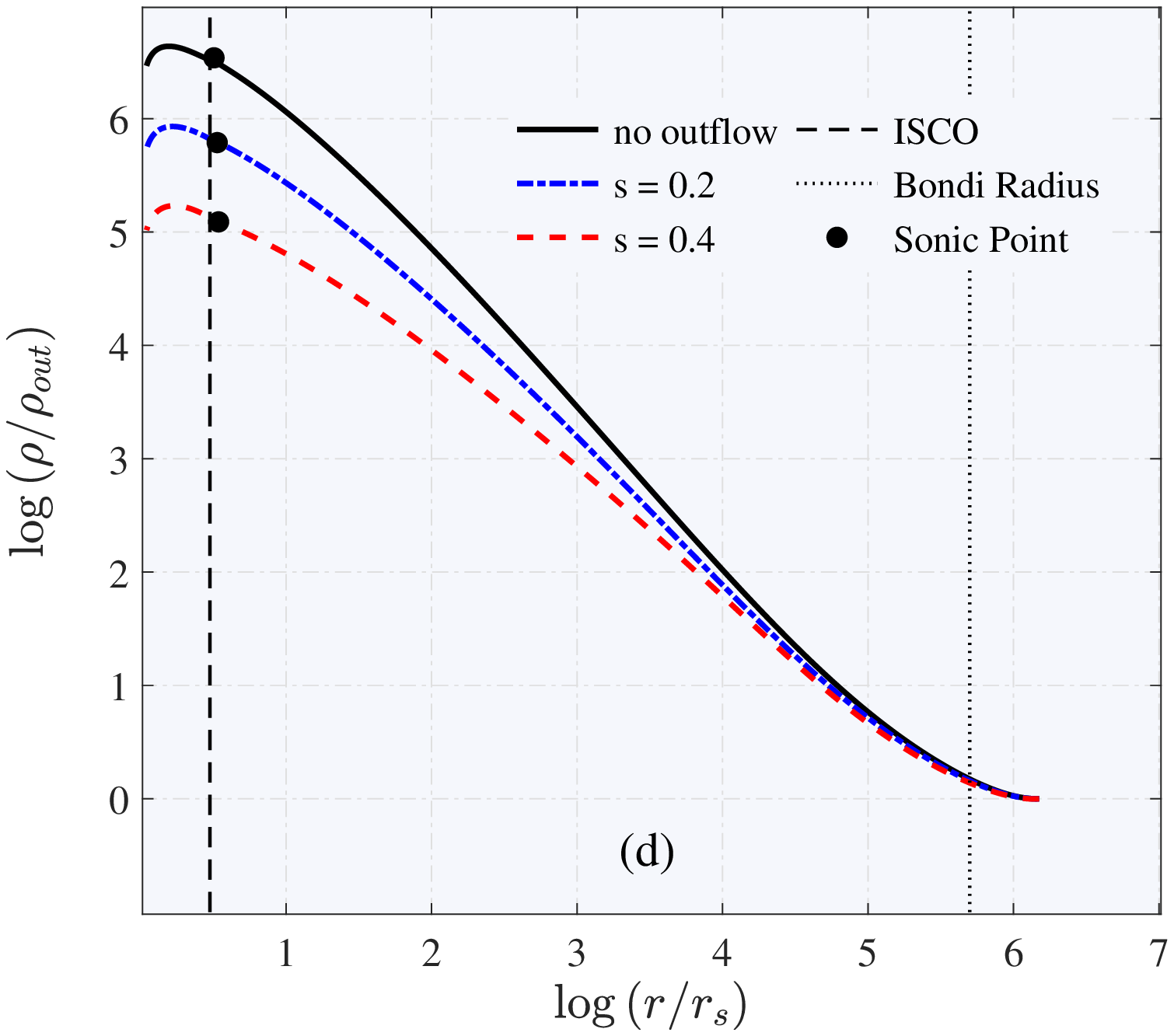}
  \end{tabular}%
  }
  \centering  
\caption{Global solutions for  $ L = 85 $, $ \alpha = 0.1 $, $ \gamma = 5/3 $ and $ T_{\rm out} = 6.5 \times 10^6 K $. The three solutions shown have ($ \rm s $, $ r_{\rm c}$) = (0, 3.1981), (0.2, 3.3651), and (0.4, 3.4381), respectively. The position of the critical point is indicated by a solid dot (saddle). The dotted curves show Keplerian angular momentum $ \Omega_{\rm K} $ and the Keplerian specific angular momentum $ l_{\rm K} = \Omega_{\rm K} r^2 $. Variations of flow variables with radius $ r $ are seen in four panels. Panel (a): The radial velocity of the accretion flow varies with radius. Panel (b): The radial variation of the angular momentum $ \Omega $ in the three solutions. Panel (c): The specific angular momentum of the accretion flow $ l_{\rm flow} = \Omega r^2 $ as a function of the radius. As a comparison, we also show the Keplerian angular momentum in the same figure. Panel (d) shows variations of the density of the flow $ \rho $. These solutions are drawn with different values of $ s $, $ s = 0 $ ( black), $ s = 0.2 $ (blue) and $ s = 0.4 $ (red). The vertical dashed line and the vertical dotted line correspond to the location of the marginally stable orbit (ISCO) and the Bondi radius $ r_{\rm B} $, respectively.}
\label{4figure} 
\end{figure*}

There is no doubt that outflow causes the surrounding environment to heat up and raise its temperature, and consequently, it makes the problems more complicated. For simplification, we assume the temperature at the outer boundary remains constant without incorporating outflow effects. The temperature of the halo surrounding most large elliptical galaxies ranges between $ 1-10 $ million Kelvin (It is in the range of the virial temperature) (\cite{2014PhRvD..90d4010R}). Indeed, the present work is an attempt to provide a more thorough survey of solutions by adopting different values of temperature in the outer boundaries. As suggested by \url{NF11}, our solutions are based on $ T_{\mathrm{out}} = 6.5 \times 10^{6} K$. According to the intended temperature at the external medium, the speed of sound is equal to $ c_{\mathrm{out}} = 10^{-3} c $, and the density contractually is considered $ \rho_{\mathrm{out}} = 1  $.

The last boundary condition that we need comes from the angular velocity in the external medium. We assume slowly rotating gas has constant angular velocity in the outer boundary. It has been assumed the outflow has no effect on surrounding gas.

\begin{equation}
\Omega = \Omega_\mathrm{out}, \quad  r = r_\mathrm{out}
\end{equation}

The angular momentum of gas is one of the most important physics quantities in rotating accretion flows. There is an open question regarding the amount and distribution of angular momentum in the accreting matter, but it is believed models with low angular momentum content are the most promising (\cite{2002A&A...383..854Y}, \cite{2007ragt.meet...35C}, \cite{2003ApJ...582...69P}). We introduce a dimensionless special angular momentum equal to the ratio of the specific angular momentum of the external gas $ l_\mathrm{out} $ to the specific angular momentum of the marginally stable orbit $ l_\mathrm{ms} $. 

\begin{equation}
L = \frac{l_\mathrm{out}}{l_\mathrm{ms}}
\end{equation}

For the Paczynsky \& Wiita potential $  l_\mathrm{ms} $ is 

\begin{equation}
l_\mathrm{ms} = \sqrt{\dfrac{27}{8}} c r_g
\end{equation}

\section{NUMERICAL RESULTS}\label{3}

In this section, we describe the numerical results of the
slowly rotating accretion flows with outflow. The quasi-spherical accretion flow with outflow drastically changes the flow structure relative to the classical Bondi solution, and the mass accretion rate onto the BH gets significantly reduced. The conventional value of $\gamma = 5/3$ is adopted in all our calculations. Also, all calculations are done with the viscosity parameter $\alpha = 0.1$, which is roughly in accordance with what is suggested by the numerical simulations \cite{2013ApJ...767...30B}, \cite{2019MNRAS.484.1724B}. We solved the main set of equations numerically using the relaxation method. The relaxation method is a powerful technique for differential equations with appropriate boundary conditions at the outer radius and the sonic point. We can obtain the global structure of the slowly rotating accretion flow in the presence of outflow by integrating these differential equations from the outer boundary to the black hole horizon. The flow passes smoothly from the sonic point and falls into the black hole (for more details, see \url{R22}). The outer radius of the accretion flow $r_{\rm out} = 3 r_{B}$ is adopted. In the inner region, a set of algebraic equations describe the flow, which is easily solved. As discussed in Section \ref{Model}, there is a set of inputs such as $s$, $\eta $, and $l$, which describe outflow properties. Mass, energy, and angular momentum can be taken away by outflows, which generally reduces the mass accretion rate.

%%\textbf{!!!!!!!Due to the fact that the mass-loss mechanisms from this system remain unknown, so thermally-driven and centrifugally-driven %%outflows are considered possible causes.!!!!! We need to change this part and put them somewhere else!!!!!!!!!} {\color{magenta}OK, I do it}

In Fig. \ref{4figure}, the global solutions for the slowly rotating
accretion flow have been illustrated by adopting various values of the mass loss power-law index $s$. These figures show the distributions of
radial velocity, angular velocity, specific angular momentum, and density, respectively. 
In panel (a) of Fig. \ref{4figure}, the radial velocity is plotted as a function of the radius with different values of $s$  while $l = 1$ and $\eta = 1$. For comparison, the case without outflows has been plotted with solid black lines. The sonic radius $r_{\rm c}$, shown by the black dots, is located almost close to the marginally stable orbit $ r_{\rm ms} = 3 r_{\rm g}$. Physically, the outflows are restrained in the inner regions by a strong gravitational force of the central black hole. This figure illustrates well how outflow affects
the outer region. Panel (b) illustrates the variation of angular velocity as a function of radius. The sloping dotted line in panel (b) shows the Keplerian angular velocity $\Omega_{K}$ while the solution with $ s = 0 $ is plotted as black lines. Panel (c) of Fig. \ref{4figure} shows the radial variation of the specific angular momentum of the accretion flow $l_{\rm flow}$. The distribution of the Keplerian angular momentum $ l_K = \Omega_K r^2 $ is shown by the dotted line for comparison. Panel (d) shows the profile of density $\rho$ for three solutions. Due to the fact that the mass accretion rate decreases with radius, the density profile of the accretion flow becomes flatter than when the mass accretion rate is constant. Consequently, the density profile flattens compared to the work of \url{NF11}. These results have been strongly supported by observations in the case of Sgr A*(\cite{2003ApJ...598..301Y}), the low-luminosity AGN NGC 3115 (\cite{2011ApJ...736L..23W}), and black holes in elliptical galaxies (\cite{2000MNRAS.311..507D}).
\begin{figure}
\centering
    \includegraphics[width=85mm]{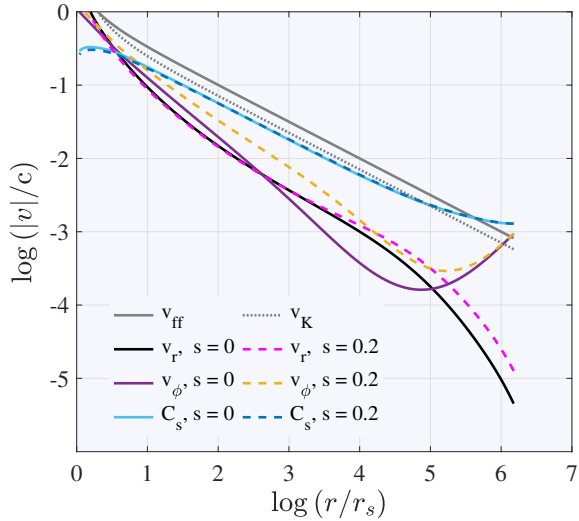}
\caption{The various velocities of the accretion flow. The adiabatic index $ \gamma = 5/3 $ is adopted in the calculations. For comparison, the Keplerian velocity $v_{\rm K}$ and free fall velocity  $v_{\rm ff}$ are plotted on top of the curves.} 
\label{V_diff} 
\end{figure}
Unlike the old picture where the accretion rate posses a constant value, many hydrodynamics (HD) and magnetohydrodynamics (MHD) numerical simulations demonstrate clearly that the inflow accretion rate decreases with decreasing radius (\cite{1999MNRAS.310.1002S}; \cite{1999MNRAS.303..309I}; \cite{2012ApJ...761..130Y}; \cite{2012ApJ...761..129Y}; \cite{2016MNRAS.455.3381S}). It is commonly accepted that this varying inflow rate is caused by a mass loss in a wind/outflow. In Figure \ref{V_diff}, we plot the velocities of the accretion flows as functions of the radius with different values of the $s$. It clearly shows that the radial velocity of the gas with outflow in outer regions is significantly higher than for the case without outflow. According to Fig. \ref{V_diff}, the radial and azimuthal velocities are different when outflow is present or absent, especially in the subsonic region, where the outflow is more effective. The radial velocity in the inner part of the flow is not affected by the outflow, but the outer part is enhanced by the rotating outflow in comparison with the no-outflow case (\cite{2013ApJ...765...96A}). For a given set of parameters, the azimuthal velocity is fairly sub-Keplerian. As a comparison, we also plot the results without outflows in the same figures. 
The azimuthal velocity profiles are plotted as a function of the radius (solid violet line ($s = 0$) and yellow dashed line ($s = 0.2$)). It is clear if the exponent s decreases, the flow will rotate more slowly than without outflows.
In Fig. \ref{Mdot}, the mass accretion rates as functions of radius are plotted
for the global solutions for various values of exponent $s$. We
compare our modified solutions with the global solution of the slowly rotating accretion flows without outflow. Generally, the mass accretion rate decreases everywhere in the flow in the presence of the outflow. It is shown that the mass accretion rate decreases towards the black hole, which is a power-law form at the outer region, while it is constant in the inner region of flow, close to the black hole.
\begin{figure}
\centering
\includegraphics[width=85mm]{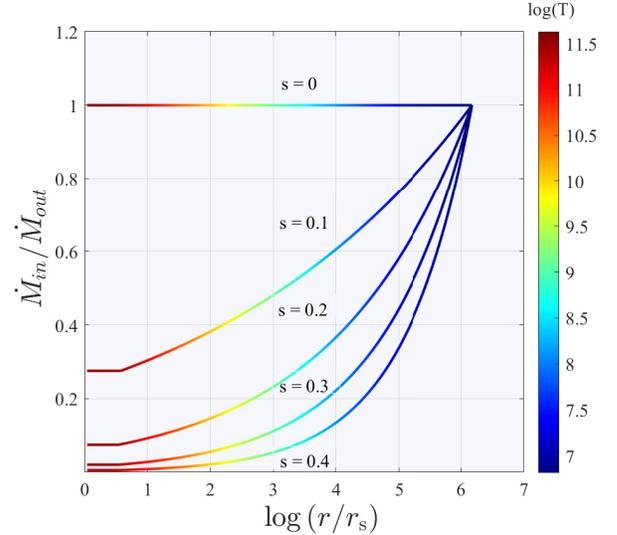}
\caption{ The mass-accretion rates of the flow as a function of radius, where $ \dot{M}_{\rm out} $ is the accretion rate at radius $ r = r_{\rm out} $. The color lines indicate the results with different values of $ s $. The top horizontal line shows the mass accretion rate without outflow ($ s = 0 $).}
\label{Mdot}
\end{figure}
Due to strong gravity near the BH, the combined fluid centrifugal force and pressure gradient force cannot overcome gravity. On the basis of this argument, it has been suggested that there will be little mass loss in the inner regions; in other words, the matter can never cross the flow surface, which causes $s$ to equal $0$. As the exponent $s$ increases, the mass accretion rate onto the black hole decreases. The color bar represents the temperature of flow. The temperature of the accretion flow decreases with the increase in mass-loss rate. This phenomenon is because, in the accretion flow, a fraction of the gravitational energy is tapped to accelerate the outflow. This reduces the heating of the flow (\cite{2009MNRAS.400.1734L}). 

\begin{figure*}
\centering
    \includegraphics[width=85mm]{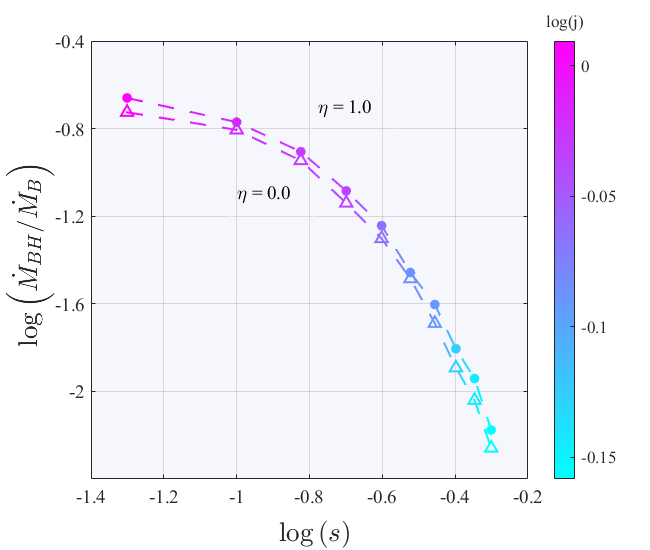}
    \includegraphics[width=85mm]{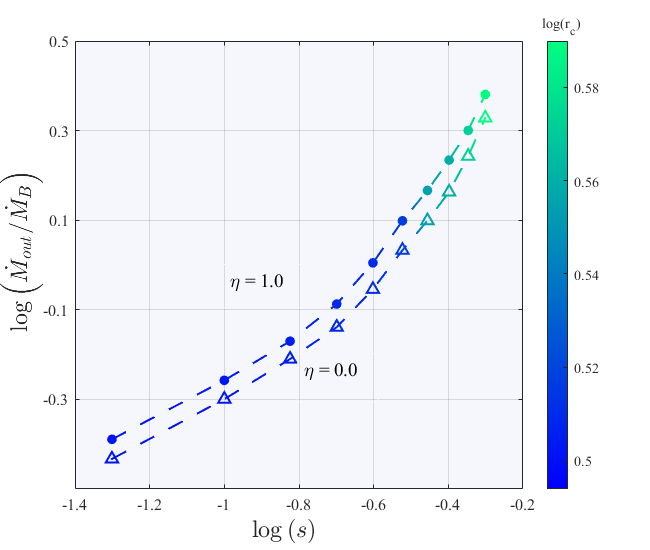}
\caption{Left panel: The mass accretion rate onto a BH as a function of exponent $s$ with different values of $ \eta $ are plotted for $ L = 85 $, $ l = 1 $ $ \alpha = 0.1 $, $ \gamma = 5/3 $ and $ T_{\rm out} = 6.5 \times 10^6 K $. The color bar shows specific angular momentum that is swallowed by a black hole. Right panel: The mass accretion rate at the outer boundary as a function of exponent $s$. The color bar shows the location of the sonic point.} 
\label{eta} 
\end{figure*}

\begin{figure*}
\centering
    \includegraphics[width=80mm]{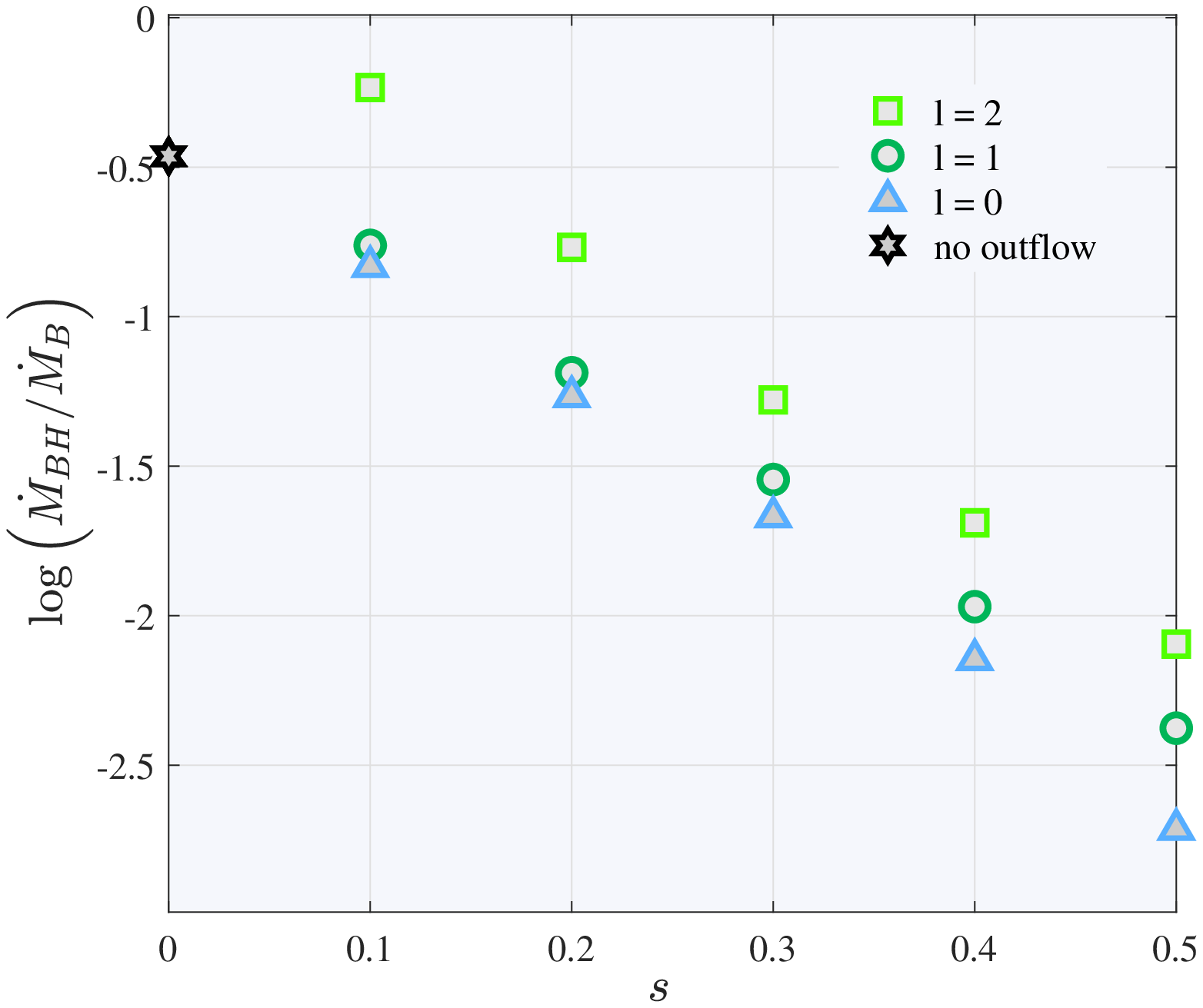}
    \includegraphics[width=80mm]{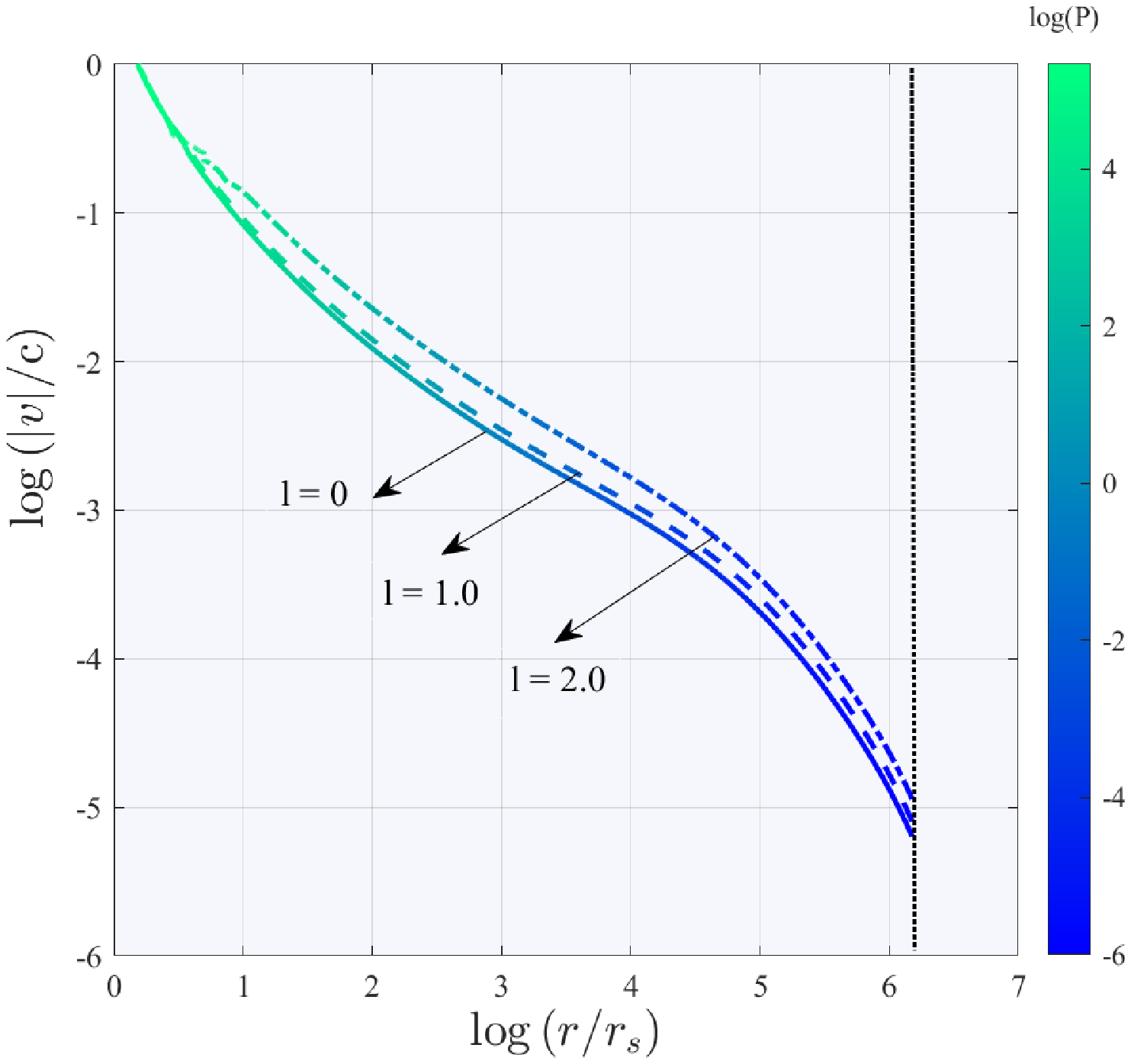}
\caption{The left panel shows the mass accretion rate onto the black hole in the unit of Bondi rate as a function of parameter $ s $. The light green square, the dark green circle, and the blue triangles represent the mass accretion rate onto the black hole for $ l = 2 $, $ l = 1 $, and $ l = 0 $, respectively. The black hexagram shows the value of the mass accretion rate in the absence of outflow. The adiabatic index $ \gamma = 5/3 $ is adopted in the calculations. The right panel shows the variation of radial velocity for three solutions $ l = 2 $, $ l = 1 $, and $ l = 0 $. The color bar shows the pressure of accretion flow.} 
\label{MdotV_l} 
\end{figure*}

\begin{figure*}
\centering
    \includegraphics[width=85mm]{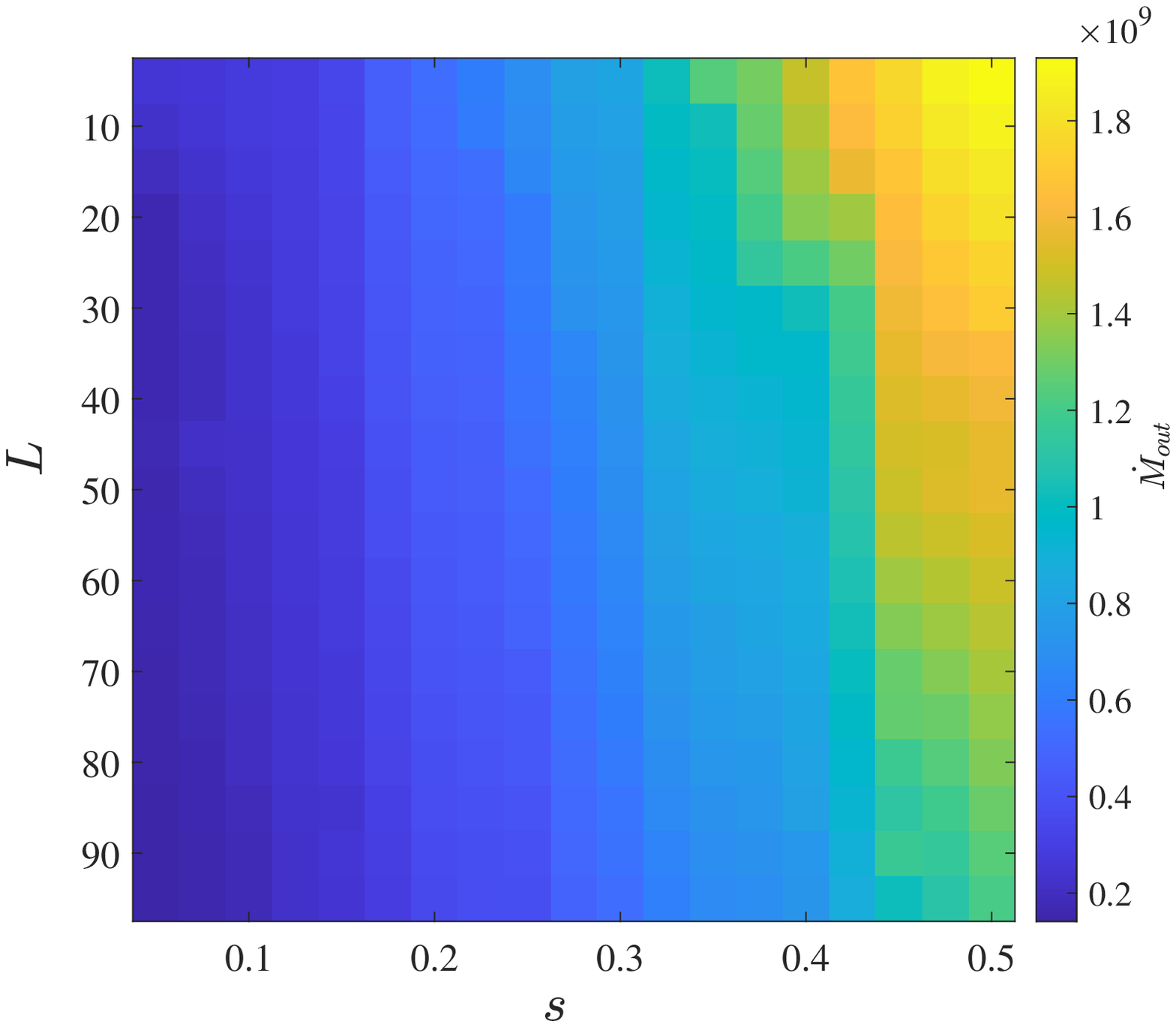}
    \includegraphics[width=85mm]{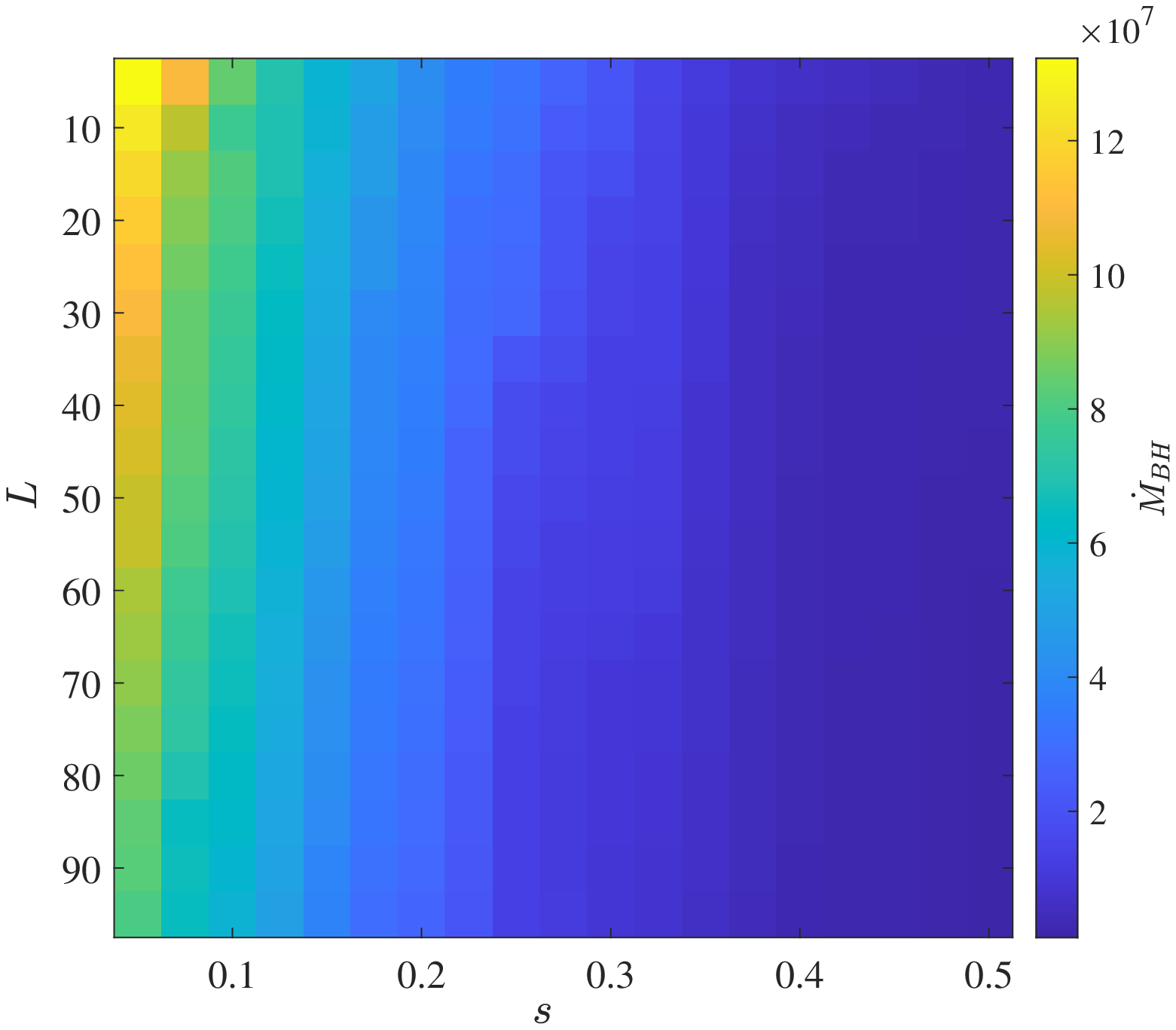}
\caption{The mass accretion rate as a function of parameters $ s $ and $ L $. Left panel: The mass accretion rate in the outer boundary $\dot{M}_{\rm out}$,
Right panel: The mass accretion rate onto the black hole $\dot{M}_{\rm BH}$. Both of mass accretion rate are in our units where $ r_g = c = \rho_{out} = 1 $. In each panel, the horizontal axis displays the strength of outflow, the vertical axis shows the angular momentum at the outer boundary, and the color bar represents the mass accretion rate. The darker shaded areas have a lower mass accretion rate, and the lighter colors have larger mass accretion rates.}
\label{3D} 
\end{figure*}

Figure \ref{eta} shows the variation of the mass accretion rate as a function of exponent $s$, for two various values of $\eta $, which represents the importance of energy carried away by outflow. Our analysis explores two extreme scenarios for the efficiency parameter $\eta$. When $\eta$ is set to 0, the impact of energy outflow on the flow is minimal, making this model suitable for situations where the outflow is primarily powered by external energy sources. On the other hand, when $\eta$ is set to 1, the effect of energy outflow on the flow is maximized, representing a model where the flow alone powers the outflow energetically, as described in \url{K99}. Additionally, the figure presented in our analysis is based on $l = 1$. For values of $l$ less than 1, even more energy can be dissipated by the disk due to the outflowing material carrying less rotational kinetic energy than it had at the point of ejection, as noted in \url{K99}. The left panel represents the BH accretion rate as a function of exponent $ s $, it well is shown mass accretion rate reduces as the intensity of mass outflow ($s$) increases. The color bar shows the specific angular momentum that the black hole swallows. It is evident that as outflow intensity increases, specific angular momentum swallowed by BH decreases slightly. The right panel shows the mass accretion rate in the outer boundary $ \dot{M}_{\rm out} $ as a function of exponent $s$. As the intensity of mass outflow ($s$) increases, the mass accretion rate in the outer boundary increases. The color bar shows the location of the sonic point. It can be well seen that the radius of the sonic point increases slightly as outflow intensity ($s$) increases. When there is outflow, the mass accretion rates at the outer boundary are higher than in the absence of outflow for the same gas conditions, including the gas angular momentum. The reason for this is that outflow removes angular momentum, so its presence leads to decreased gas angular momentum and, therefore, a higher mass accretion rate at the outer boundary \cite{2018EPJWC.16804005P}. Furthermore, as can be seen in the figure, the mass accretion rate increases by considering outflow cooling effects. In this figure, we see different $\dot{M}$ for two value of $\eta$ parameter ($\eta = 1$ and $\eta = 0$). There is only a minimal difference between when the outflow carries energy ($\eta = 1$) and when it does not ($\eta = 0$).

In the following, we tend to focus on the role of the dimensionless lever arm $l$. Figure \ref{MdotV_l} shows the variation of mass accretion rate (left panel) and radial velocity (right panel) for different $l$. In the left panel, it is well seen that the mass accretion rate onto the black hole is greater for $ l = 2 $ than $ l = 0$. Considering that for the case $l = 0$, the flow loses only mass, which corresponds to a non-rotating outflow. In the case of $ l = 2 $, the outflows carry away a considerable amount of the angular momentum of the flow and cause its structure to become significantly different from the conventional viscous flow structure. In the right panel, the radial velocity for $l = 2$ is larger than $ l = 0 $. Despite the fact that the radial velocity in the innermost region of the flow doesn't change due to the outflow, a rotating outflow increases the radial velocity in the outer region of the flow. According to this study, the radial velocity of an accretion flow in which the angular momentum is removed by the outflows is slightly higher than that of an accretion flow in which the angular momentum is not removed by outflow. In this panel, the color bar presents the pressure of the flow as a function of radius. It was assumed that outflow material would rotate with the flow. The non-rotating outflow corresponds to $ l = 0 $. As $ l $ exceeds $0$ ($ l>0 $), the angular momentum is carried away by the outflow. In addition, outflows can remove more angular momentum from the flow, corresponding to $ l > 1 $, such as the centrifugally driven magnetohydrodynamic winds (\cite{2022ApJ...930..108W}, \cite{1982MNRAS.199..883B}).

Figure \ref{3D} represents the result of the solution for $ s = 0.05-0.5 $ and $ L = 5-95 $. Each panel of Figure \ref{3D} displays an image that uses the full range of colors in the colormap. Each element specifies the color for one pixel of the image. The resulting image is a 19-by-19 grid of pixels where 19 is the number of rows, and 19 is the number of columns. The row and column indices of the elements determine the centers of the corresponding pixels. In each panel, the color bar shows the mass accretion rate. The mass accretion rates are in our units where $ r_g = c = \rho_{out} = 1 $. Areas with darker colors have lower mass accretion rates, and lighter colors have larger mass accretion rates. It is clear that the lower right region of each panel, which represents the larger $ s $, i.e., the more strength outflow, is the Disk-like type (larger $ L $). In contrast, the upper left region in each panel indicates the smaller $ s $, i.e., weaker outflow, is the Bondi type one (smaller $ L $). According to the right-hand panel in Fig.\ref{3D}, $\dot{M}_{BH}$ decreases as the rotation of the external gas ($L$) increases. The left panel shows the variation of mass accretion rate in the outer boundary as a function of parameters $ s $ and $ L $. It is clear with increasing the $ s $, the mass accretion rate in the outer boundary increases. The right panel shows the variation of mass accretion rate in the black hole as a function of parameters $ s $ and $ L $. It is clear that as exponent $ s $ increases, the mass accretion rate onto BH decreases, indicating that outflow is taking away mass. 

To describe Table \ref{table}, we use two subclasses of the
model. The models in Table \ref{table} are classified into two
subclasses, i.e., models A1-A3 and B1-B3. Here,
models A1-A3 are called A-type models, while models
B1-B3 are called B-type models. The outer boundary of our calculation is $ 3 r_{\rm B}$ for the A-type
models while $ r_{\rm B} $ for the B-type models. Our numerical solutions with different outer boundaries are summarized in Table \ref{table}. Columns 2 and 3 of Table \ref{table} are shown the temperature and angular momentum at the outer boundary. Columns 4-6 include outflow parameters, $s$, $\eta$, and $L$, respectively.

 Column 7 shows the location of the outer boundary in units of the Bondi radius, and column 8 shows the galaxy potential. Columns 9 and 10 show the mass accretion rate onto BH and at the outer boundary in units of Eddington accretion rate, respectively. We compare models with different outer boundary conditions,
i.e. models A1, A2, and B1, B2, and give results in Tabel \ref{table}. The results show that compared to model A2, whose BH accretion rate is smaller than that of model A1, both exponent $s$ and specific angular momentum at the outer boundary $L$ in model A2 have a higher value. In models B1 and B2, the temperature at the outer boundary is larger than A1 and A2, and the BH accretion rate is less. Also, we compare the solutions of models A2 and A3. Model A2 includes the galaxy potential, while model A3 does not include it. It can be seen in Tabel \ref{table} that the galaxy potential increases the BH accretion rate, slightly as had been achieved in our previous paper (\url{R22}). The same result can be obtained from the comparison of the B1 and B2 models. In essence, the present results show that galactic potential may play a significant role in the dynamics of the slowly rotating accretion flows and their outflow in massive ellipticals. Therefore, it is important not to overlook the galactic potential when modeling realistic (numerical and analytical) accretion dynamics. As a result, this would provide a better understanding of the energy of AGN feedback in the context of these massive galaxies in the contemporary universe. 

\begin{table*}
\caption{Summary of Our Model} % title name of the table  
\centering % centering table  
\begin{tabular}{l c c rrrrrrr} % creating 10 columns  
\hline\hline   
 Model ID & $T_{\rm out}$ & $ L $ & $ s $ & $\eta$ & $l$ & $r_{\rm out}$ & $\phi_{\rm galaxy}$ & $\dot{M}_{\rm BH}$ &  $\dot{M}_{\rm out}$
\\ [1ex]      & $(k)$ &  & & & & $(r_{\rm B})$ & & ($\dot{M}_{\rm Edd}$) & ($\dot{M}_{\rm Edd}$) 
\\ [1ex] 
\hline   
% Entering 1st row  
 A1 & 3.9$\times$ $10^6$ &  5 & 0.2 & 1 & 1 & 3 & ON & $1.11 \times 10^{-5}$ & $1.62 \times 10^{-4}$ \\[-1ex]  
% Entering 2nd row  
\\[1ex]  A2 & 3.9$\times$ $10^6$ &   85 & 0.5 & 1 & 1 & 3 & ON & $6.73 \times 10^{-7}$ & $5.68 \times 10^{-4}$ \\[-1ex]  
% Entering 3rd row  
\\[1ex]  A3 & 3.9$\times$ $10^6$ &   85 & 0.5 & 1 & 1 & 3 & OFF & $4.46 \times 10^{-7}$ & $3.76 \times 10^{-4}$ 
 
% [1ex] adds vertical space 
\\[1ex] 
\hline
% Entering 1st row
  B1 & 6.5$\times$ $10^6$ &  5 & 0.2 & 1 & 1 & 1 & ON & $7.64 \times 10^{-6}$ &  $8.10 \times 10^{-5}$ \\[-1ex]  
% Entering 2nd row  
\\[1ex] B2 & 6.5$\times$ $10^6$ &  85 & 0.5 & 1 & 1 & 1 & ON & $4.89 \times 10^{-7}$ & $1.7 \times 10^{-4}$ \\[-1ex]  
% Entering 3rd row  
\\[1ex] B3 & 6.5$\times$ $10^6$ &  85 & 0.2 & 1 & 1 & 1 & OFF & $3.48 \times 10^{-7}$ & $1.32 \times 10^{-4}$ 

 \\[0.5ex]
\hline
\hline % inserts single-line
\end{tabular}
 \\[1ex]
\footnotesize{Note. Column (2) is the temperature in the outer boundary; Column (3) is the angular momentum at the outer boundary,
 Columns 4-6 are parameters of outflow; column (7) shows the location of the outer boundary; column (8) is the galaxy potential and Column (9)
and (10) the mass accretion rate in the black hole and the outer boundary, respectively.}
\label{table}
\end{table*}

\subsection{Comparison with previous works} \label{thickness}

The purpose of this section is to compare the results with the simulation performed by \cite{2019MNRAS.484.1724B} hereafter\url{BY19} and the solution presented in our previous paper, \url{R22}. In order to compare our model with them properly, we consider the black hole mass equal to $10^8 M_{\odot}$  ($M_{\odot}$ is solar mass) and the density at the outer boundary $\rho_0$ $ = 10^{-24}$ $gr$ $cm^{-3}$. Figure \ref{simu} shows the mass accretion rate of the black hole in the unit of $ \dot{M}_E $ as a function of temperature at the outer boundary. The dashed line shows the Bondi accretion rate by assuming $\gamma = 5/3$. We present our numerical solutions for three different values of exponent $s$. The blue circles show our numerical solution without considering the outflow that corresponds to $ s = 0 $. The pink diamonds are our solution when $ s = 0.1 $. Solutions with $ s = 0.3 $ can be seen as green squares. The orange squares are according to the time-averaged values of the mass accretion rate of simulations in \url{BY19}, and the error bars on squares correspond to the change range of simulations due to fluctuations. The solid blue line is from the fitting formula (Equation (5) of \url{BY19}). The fitting formula is an analytical formula that can accurately predict the luminosity of observed LLAGNs (with black hole mass $\sim 10^8 M_{\odot}$). The black hole accretion rate is calculated using this formula based on the density and temperature of the gas at the parsec scale. It is clear that the BH accretion rate of solutions without outflow is larger than that of other solutions. It well has shown outflow can reduce the mass accretion rate onto the black hole. We find that the solution corresponding to $ s = 0.3 $ can represent the fitting formula well. Furthermore, we mention that they considered $ s = 0.5 $. Compared with \url{YB19} simulations, our results deviate slightly due to ignoring radiation effects. Perhaps it is because our results agree with simulation results at $ s = 0.3 $. 

In the last step, by keeping the assumption of vertical hydrostatic equilibrium similar to our previous work, \url{R22}, we investigate the effect of outflow on the thickness of flow. In this part, we assume the initial state $ H \ne r $ rather than the simple approximation $ H = r $. Fig.\ref{H} indicates the radial variation of the relative thickness $ H / r $ of the flow in the presence of outflow for various values of exponent $ s = 0, 0.2, 0.4 $. The color bar shows the density of accretion flow. It has been demonstrated that flow without outflow is thicker than flow in the presence of outflow. It has been well seen that outflow reduces the thickness of the flow. In other words, the thickness of the flow increases as the intensity of the outflow decreases. As is evident from the results, at intermediate radii, such as ($10 - 10^5 $) $r_{\rm s}$, the flow thickness $ H/r $ can vary between $0.4$ to $0.6$ for different $ s $ ($ 0.4 - 0 $). According to our results, accretion flow with stronger outflows will be thinner than those without. In this part, the outer boundary is fixed in the Bondi radius $r_{\rm out} = r_{\rm B}$, similar to our previous work (\url{R22}).

\subsection{Mechanical AGN Feedback}\label{sec.AGN}

In this section, we discuss in detail how the properties of outflow are calculated for a given accretion rate. 
We want to know, for a given accretion rate, what will be the output of AGN, or what will wind properties be?
In our accretion mode, accretion flow produces outflow but not radiation and jet. It is assumed that jets deposit very little energy in the galaxy, so we neglect it here. In the future, this assumption needs to be examined. As the wind is launched from the accretion flow, it will affect the black hole's accretion rate. A fraction of the gas in the accretion flow is ejected as a form of wind. So the outflow mass flux can be obtained from Eq. \ref{wind}.

As a result, the accretion rate close to the horizon of a black hole, which determines its luminosity, is described by the following equation:

\begin{equation}
\dot{M}_{\rm BH} = \dot{M}_{\rm out} (\frac{r_{\rm c}}{r_{\rm out}})^s
\end{equation}

Similar to \cite{2018ApJ...857..121Y} work, we assume $ s = 0.5 $ in this section. The opening angle of the outflow is greater than that of jets, which makes the interaction between the wind and the ISM more efficient (refer to Fig. 1 in \cite{2015ApJ...804..101Y}). According to previous investigations (\cite{2015ApJ...804..101Y}, Equation (8)), outflow's poloidal velocity roughly remains constant and is approximated as follows:
\begin{equation}
 v_{\rm w} = (0.2-0.4) v_{\rm k}(r_{\rm out})  
\end{equation}

Here $v_{\rm k}$ is the Keplerian velocity at the outer radius $r_{\rm out}$.

Here we assume $v_w = 0.2 v_k(r_{\rm out})$ and consequently, the flux of energy and momentum of outflow will be introduced as:

\begin{figure}
\centering
\includegraphics[width=85mm]{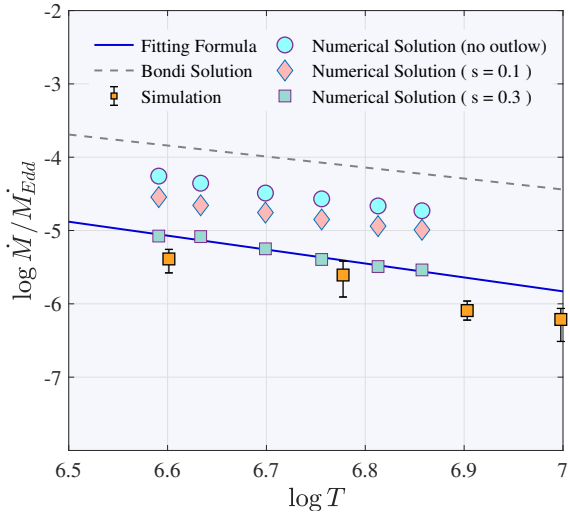}
\caption{The mass accretion rate of the black hole in the unit of $ \dot{M}_E $ as a function of temperature at the outer boundary $T_{\rm out}$. The density at the outer boundary is considered
$ \rho_0= 10^{-24} gr/cm^3 $. Solutions are for $ L = 85 $, $ \alpha = 0.1 $, $ r_{\rm out} = 3 r_{B} $ and $ \gamma = 5/3 $. 
} 
\label{simu} 
\end{figure}

\begin{equation}\label{Edot}
\dot{E}_w = \frac{1}{2} \dot{M}_w v_k^2,
\end{equation}
,
\begin{equation}\label{Pdot}
\dot{P}_w = \frac{2 \dot{E}_w}{v_w},
\end{equation}

By comparing the energy and momentum fluxes of the wind in the cold and hot feedback modes with those produced by our model, we are able to evaluate the validity of our parametric model. The results of such a comparison are shown in Fig. \ref{AGN}. The left and right panels denote the power and momentum fluxes, respectively. 
In both hot and cold modes, wind momentum flux is large, suggesting that outflow will be instrumental in pushing gas away from an AGN. As a comparison between our current description of outflow and previous studies, we have also shown the AGN model from \cite{2014ApJ...789..150G}. It can be seen from Fig. \ref{AGN} that slowly rotating accretion flows have a lower black hole accretion rate than hot and cold modes ($10^{-5}-10^{-7} \dot{M}_{Edd}$). Here, $ \dot{M}_{Edd} = 10 L_{Edd}/c^2$ is the Eddington accretion rate.

Also, the wind's energy and momentum flux is less than cold and hot feedback modes. This means that the wind is less powerful in an accretion flow that rotates slowly.
Depending on the parameter $s$, the wind power ranges from $ 10^{-10}-10^{-12} L_{Edd}$ and the momentum flux of wind ranges from $ 10^{28}-10^{30} $$gr$$ cm s^{-2}$.
As parameter $s$ increases, the energy and momentum fluxes of wind slightly increase. In addition, we showed the solutions for two values of angular momentum ( $L = 85$ and $L = 5$ ) at the outer boundary. As we can see, power and momentum fluxes are not much different in this both modes. We should not judge which is important in the feedback simply from its magnitude of power and momentum flux. Due to the different cross-sections of photon-particle and particle-particle interactions, the outflow must travel very different distances to convert its energy and momentum to the ISM. In this study, we ignore the effects of mechanical AGN feedback on the cosmological formation of elliptical galaxies and black hole growth.

\begin{figure}
    \centering
    \includegraphics[width=85mm]{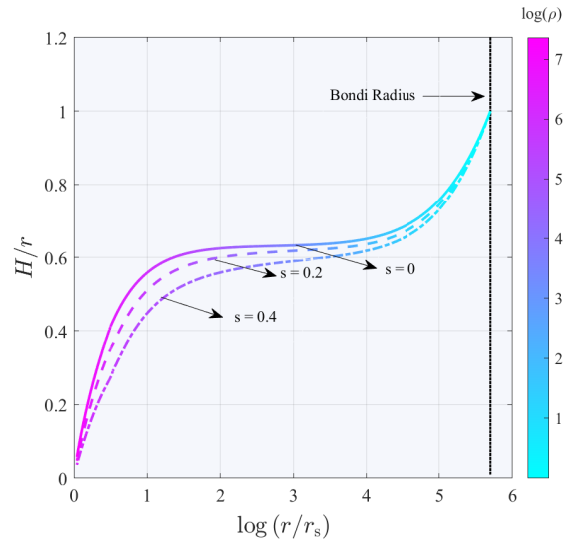}
    \caption{The radial variation of the relative thickness of the flow $ H/r $ for $ s = 0 $, $ s = 0.2 $ and $ 0.4 $. The vertical dotted line indicates the location of the Bondi radius. Solutions are for $ L = 85 $, $ \alpha = 0.1 $, $ \gamma = 5/3 $ and $ T_{\rm out} = 6.5 \times 10^6 K $, $\eta = 1 $, $ r_{\rm out} =  r_{B }$. The color bar shows the density of accretion flow.} 
    \label{H}
\end{figure}

\section{DISCUSSION AND SUMMARY} \label{sec:5}

Outflows play a significant role in the active galactic nuclei feedback process. Our aim of this paper is to investigate the influence of outflow on the structure and physical properties of slowly rotating accretion flow.
Pioneer works of numerical simulation of black hole accretion flows have found that the mass accretion rate
decreases inward (\cite{2012ApJ...761..129Y}). It can be seen that the hydrodynamic wind – that is, the outflow – helps maintain the accretion process by carrying mass, energy, and angular momentum away. The gas angular momentum or outflow could lead to a decrease in the mass accretion rate,
which will profoundly affect the growth of supermassive black holes and their coevolution with host galaxies.
The outflows from the central engine are tightly coupled with the surrounding gaseous medium, providing the dominant heat source and preventing runaway cooling. This study does not consider the interaction between the outflow and its surrounding environment. According to many previous studies, the outflow is induced by
assuming the mass accretion rate $ \dot{M} $ to be a power-law
function of radius $ \dot{M} \propto r^s $ (e. g. \cite{1999MNRAS.303L...1B}). So based on the assumption that the outflow mass-loss rate follows a power-law distribution in
radius, the calculations were simplified. The parametric approach used in the present study can be applied
to a wide range of dynamical outflow models, including radiation-driven and centrifugally-driven outflows. Similar to our previous work, a transition from subsonic to supersonic occurs as matter moves inward with increasing radial velocity along the radial direction. Our study assumes that mass accretion rate $\dot{M}$ decreases towards the black hole, which has a power-law r-dependence at larger radii but is constant in the inner region of flow close to the black hole. For $ r \leq 10 r_{\rm s} $, we considered
only the inflow region, and flow is supersonic.
\begin{figure*}
    \centering
    \includegraphics[width=85mm]{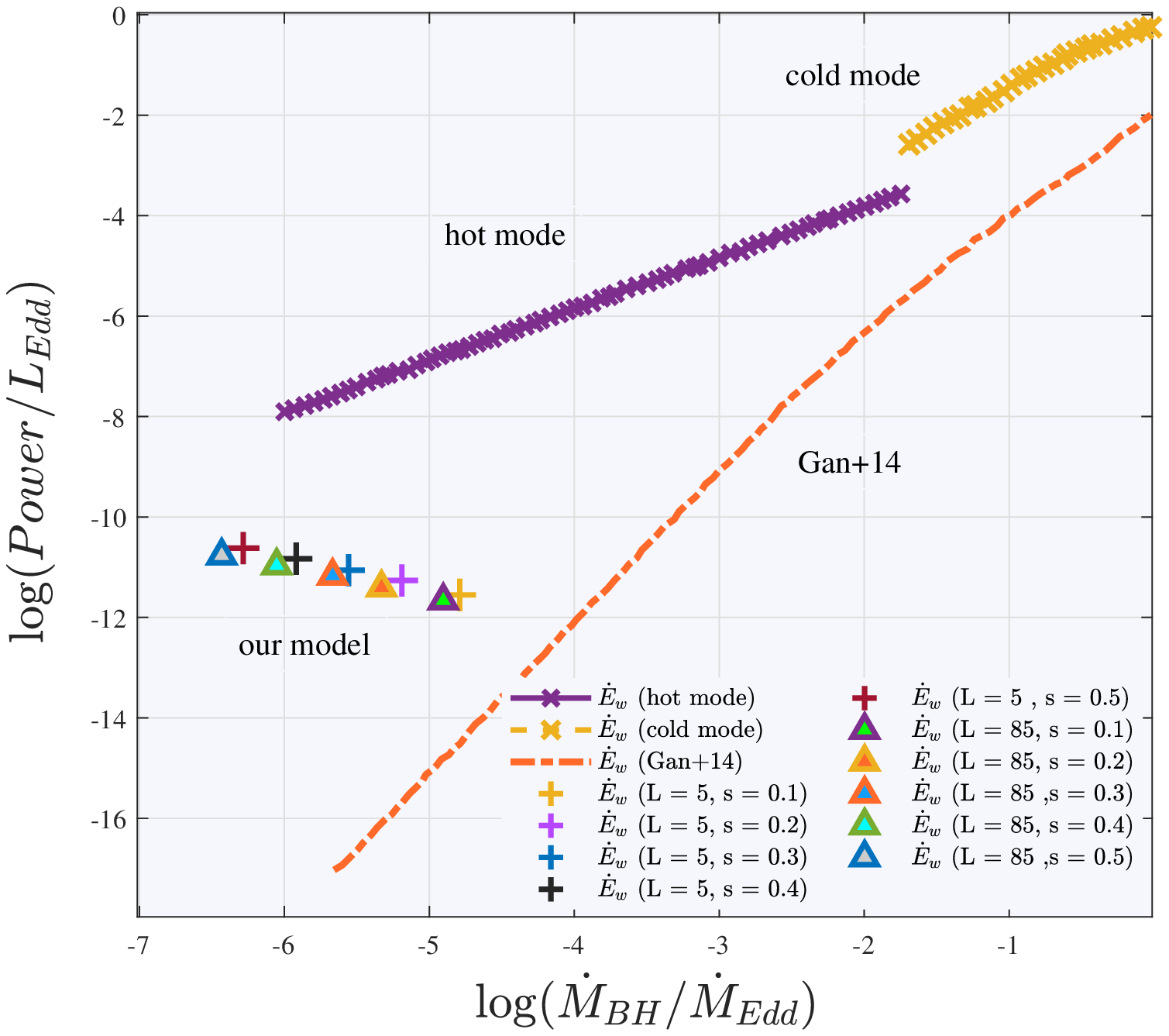}
    \includegraphics[width=85mm]{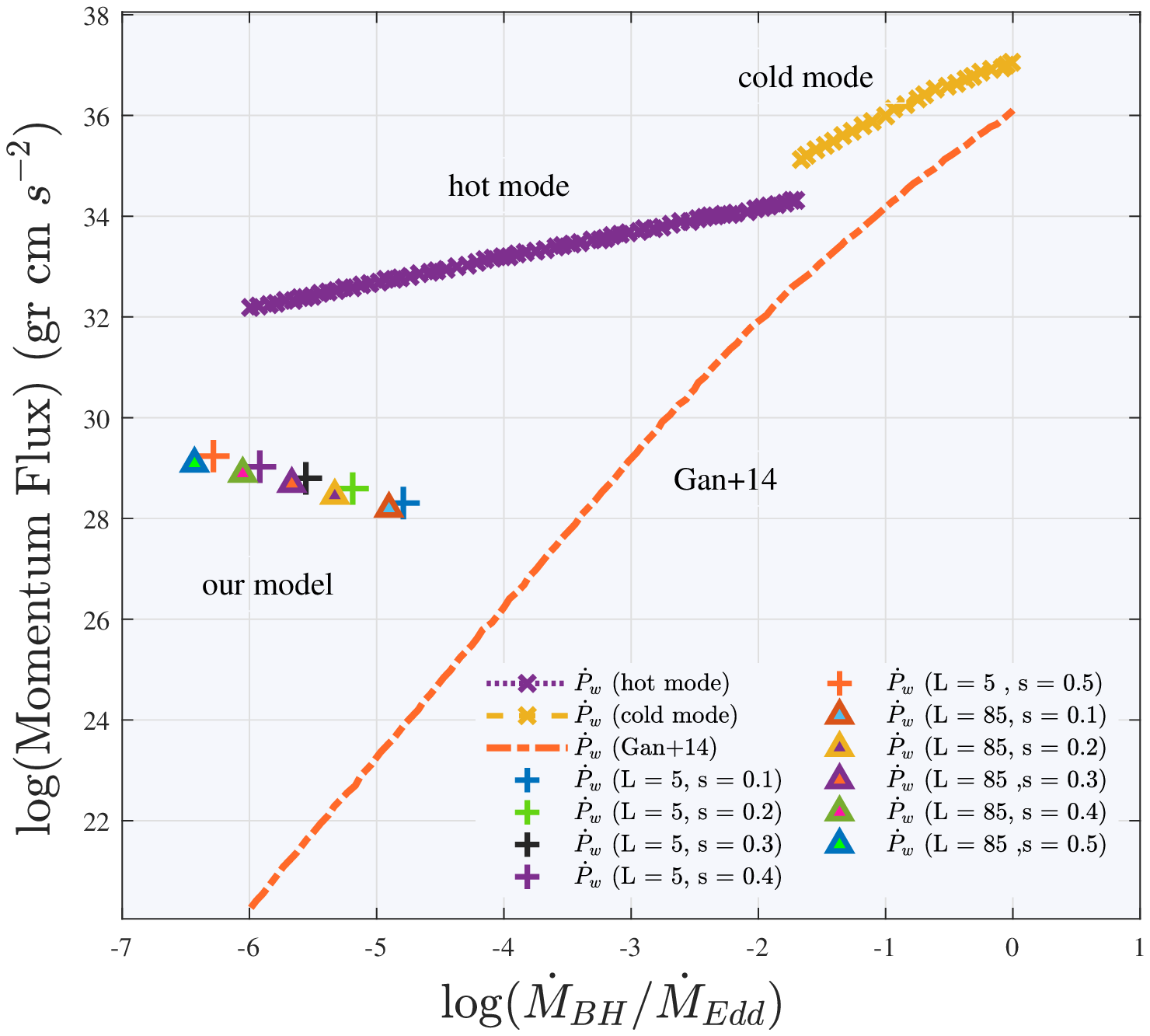}
    \caption{For $ L = 85 $, $ \alpha = 0.1 $, $ \gamma = 5/3 $ and $ T_{\rm out} = 6.5 \times 10^6 K $, $\eta = 1, $. The comparison of the power (left) and momentum flux (right) of outflow from hot (Violet lines) and cold (Orange lines) accretion modes of AGNs with our model. As a comparison, red dot-dashed lines show the wind's power and momentum flux from Gan et al. (2014).}
    \label{AGN}
\end{figure*}

%%%%%%%%%%%

We find that in the presence of outflow, the structure
of the slightly rotating accretion flow is significantly altered. 
%%We also find that the temperature of the accretion
%%flow decreases with the increase in mass-loss rate (see
%%Figs and ). The reason for this phenomenon is that in
%%the accretion flow, a fraction of the gravitational energy
%%is tapped to accelerate the outflow. This reduces the
%%heating of the flow \cite{2009MNRAS.400.1734L}. 
The outer boundary is usually located at several hundred or thousands of Schwarzschild radius ($ \sim 3 r_B \sim 10^6 r_{\rm s} $). The effect of radiative cooling on outflow strength has not been discussed in
these papers.

To be more precise, this model has modified the classical Bondi model, and the angular momentum of the gas is very small. The potential energy of the black hole, as well as the host galaxy, is taken into account. 
The main results of this work are briefly described
below 

\begin{itemize}

\item The mass accretion rate of the slowly rotating accretion flows with outflows is no longer constant radially. It has been observed that in our model, the radial density profile is flatter because the mass accretion rate decreases with decreasing radius. Basically, if the mass accretion rate were constant with the radius, density profiles would be steeper.

\item It is found that the radial velocity of the flow with winds at the outer edge of the accretion flow is significantly higher than that of a conventional accretion flow.

\item Based on our results, the flow thickness decreases as the outflow intensity increases.

\item
Our research has uncovered a relation between the s-exponent of the outflow and the ratio of the accretion rate to the Bondi accretion rate. Specifically, we have observed that as the intensity of mass outflow, denoted by s, increases, the mass accretion rate onto the black hole decreases. Conversely, an increase in s leads to a rise in the mass accretion rate at the outer boundary. 

\item Owing to the galactic contribution to the potential,
there is an enhancement in the value of mass accretion
rate relative to the initial case, and the galactic contribution of the potential has a significant effect on the
velocity profile.

\item The mass accretion rate onto BH in our solutions is in good agreement with the formula obtained of \url{YB19} that is based on the density and temperature of the gas at the parsec scale. This formula accurately predicts the luminosity of LLAGNs (with black hole mass $ \sim 10^8 M_{\odot}$).

\item The energy and momentum fluxes of outflow in our model are less than cold and hot feedback modes. AGN feedback in slowly rotating accretion flows probably has less impact on the surrounding environment due to a weak outflow.

\end{itemize}

The solutions we have developed and the flow structure differ from previous analytical studies in many
aspects (e.g., boundary conditions, the behavior of solutions, and the subsonic region). Outflows can very effectively affect the properties of the gas surrounding the AGNs (\cite{2018ApJ...857..121Y}). There is also evidence that wind/outflow can suppress star formation more strongly than radiation. 

When the jet flow reaches a particular distance from the source, it encounters the wind wall, which creates an oblique shock that changes the direction of the jet. This collision between the jet and wind also generates both $\gamma$-ray and radio flares, as shown in \cite{2023ApJ...944..187W}. In this study, we have not taken into account the interaction between winds and interstellar medium (ISM). We plan to pursue this interaction in our future work. 

\section*{ACKNOWLEDGMENTS}

We hereby acknowledge the Sci-HPC center of the Ferdowsi University of Mashhad where some of this research was performed. Also, we have made extensive use of the NASA Astrophysical Data System Abstract Service. This work was supported by the Ferdowsi University of Mashhad under grant no. 57030 (1400/11/02).



\end{document}